\pdfoutput=1
\documentclass{aa}  

\usepackage{graphicx}
\usepackage[utf8]{inputenc}

\usepackage{txfonts}
\usepackage{color}
\usepackage{lscape}
\usepackage{longtable}
\usepackage{tablefootnote}
\usepackage{booktabs}
\usepackage{hyperref}
\usepackage{subfigure}
\usepackage[normalem]{ulem}
\usepackage[dvipsnames]{xcolor}
\usepackage{tikz}
\usetikzlibrary{arrows.meta, positioning, shapes.geometric}
\usepackage{pdflscape}   
\usepackage{array}       
\usepackage{rotating}   
\usepackage{placeins}
\usepackage{nicefrac}
\newcommand{\kms}{km\,s$^{-1}$}
\newcommand{\vsini}{$\nu \sin i$}
\newcommand{\teff}{$T_{\rm eff}$}
\newcommand{\logg}{$\log\,{g}$}
\newcommand{\feh}{[Fe/H]}
\newcommand{\meh}{[M/H]}
\newcommand{\vmic}{$\nu_{\rm mic}$}

\newcommand{\mstar}{$M_{\star}$}

\newcommand{\Mj}{M$_{\rm J}$}

\begin{document} 

\title{Ariel stellar characterisation}
\subtitle{IV. Fundamental parameters of 18 hot stars in the Ariel mission candidate sample}
   \author{          
    H. Ramler\inst{\ref{TO}}\and
    S. P. D. Borthakur\inst{\ref{TO},\ref{inst2},\ref{inst3}}\and
    C. P. Folsom\inst{\ref{TO}}\and
    D. Bossini\inst{\ref{dfa},\ref{oapd}}\and
    A. Lehtmets\inst{\ref{TO}}\and
    C. Danielski\inst{\ref{unival},\ref{oaa}}\and
    D. Turrini\inst{\ref{oato}}\and
    M. Benito\inst{\ref{TO},\ref{IAC},\ref{ULL}}\and
    M. Tsantaki\inst{\ref{oaa}}\and
    L. Magrini\inst{\ref{oaa}}\and
    N. Moedas \inst{\ref{iaps}}\and
    K. Biazzo, \inst{\ref{oar}}\and
    R. da Silva, \inst{\ref{oar},\ref{asi}}\and
    M. Kama\inst{\ref{UCL}}\and 
    E. Siimon\inst{\ref{TO}}\and
    V. Mitrokhina\inst{\ref{TO}}\and        
    K. G. He{\l}miniak \inst{\ref{nicolaus}}\and 
    S. Benatti \inst{\ref{oap}}\and
    M. Rainer\inst{\ref{oabr}}
          }

   \institute{
   Tartu Observatory, University of Tartu, Observatooriumi 1, T\~{o}ravere, 61602, Estonia\\
   \email{heleri.ramler@ut.ee}\label{TO}
    \and
    Space Research Institute, Austrian Academy of Sciences, Schmiedlstrasse 6, 8042, Graz, Austria\label{inst2}
    \and
    Institute for Theoretical and Computation Physics, Graz University of Technology, Petersgasse 16, 8010 Graz, Austria\label{inst3}
    \and
    Department of Physics and Astronomy G. Galilei, University of Padova, Vicolo dell'Osservatorio 3, I-35122, Padova, Italy\label{dfa}
    \and
    INAF -- Osservatorio Astronomico di Padova, Vicolo dell'Osservatorio 5, 35122, Padova, Italy \label{oapd} 
    \and
    Department of Physics and Astronomy, University College London, Gower Street, London, WC1E 6BT, UK\label{inst4}
    \and 
    Departament d'Astronomia i Astrof\'{i}sica, Universitat de Val\`{e}ncia, Av. Vicent Andrés Estellés 19, 46100, Burjassot, Spain z\label{unival}
    \and
    INAF -- Osservatorio Astrofisico di Arcetri, Largo E. Fermi 5, 50125 Firenze, Italy\label{oaa}
    \and
    INAF -- Osservatorio Astrofisico di Torino, Via Osservatorio 20, 10020 Pino Torinese, Italy\label{oato}
    \and
    Instituto de Astrofísica de Canarias, Calle Vía Láctea s/n E-38206 La Laguna, Santa Cruz de Tenerife, España\label{IAC}
    \and
    Universidad de La Laguna, Avda. Astrofísico Francisco Sánchez E-38205 La Laguna, Santa Cruz de Tenerife, España\label{ULL}
    \and
    Institute for Space Astrophysics and Planetology INAF-IAPS, Roma, RM, Italy\label{iaps}
    \and 
    INAF -- Osservatorio Astronomico di Roma, Via Frascati 33, 00040 Monte Porzio Catone (RM), Italy\label{oar}
    \and
    Agenzia Spaziale Italiana, Space Science Data Center, via del Politecnico snc, 00133 Rome, Italy\label{asi}
    \and 
        Department of Physics and Astronomy, University College London, Gower Street, London, WC1E 6BT, UK\label{UCL}
    \and
    INAF -- Osservatorio Astronomico di Palermo, Piazza del Parlamento, 1, 90134 Palermo, Italy\label{oap}
    \and
   INAF -- Osservatorio Astronomico di Brera, Via E. Bianchi 46, 23807 Merate (LC), Italy \label{oabr}
    \and
   Nicolaus Copernicus Astronomical Center, Polish Academy of Sciences, ul. Rabia\'{n}ska 8, 87-100 Toru\'{n}, Poland \label{nicolaus}
}

  \abstract
   {The characterisation of exoplanetary systems depends on the accurate determination of host star parameters. The Ariel mission will probe the atmospheres of a statistically significant sample of exoplanets, and so requires a precise characterisation of the stellar properties well before its launch in 2029. The homogeneous determination of stellar parameters for Ariel will enable both the optimisation of the final target list and set roots for a reliable interpretation of the formation and evolution of planetary systems. Such a homogeneous characterisation has thus far only been carried out for the cool (\teff\ $\lesssim 7000\,$K) host stars among the Ariel target candidates.}
   {We present a uniform determination of fundamental stellar parameters for $18$ hot stars ( \teff\ $\gtrsim$  $7000$\,K) in the Tier 1 candidate list of the Ariel mission candidate sample.}
   {We adopted an iterative spectro-trigonometric approach optimised for high-temperature stars ( \teff\ $\gtrsim$  $7000$\,K). High-resolution spectra were analysed using the \textsc{zeeman} code with $\chi^2$ minimisation, combining model fits to metal and Balmer lines. Surface gravity was refined using photometry-based radii and masses from stellar evolutionary tracks.}
   {We derived effective temperatures, surface gravities, projected rotational velocities, microturbulent velocities, overall metallicities, iron abundances, stellar masses, and radii for our sample of $18$ hot stars. Our results were validated against a set of benchmark stars previously presented in the literature, confirming that our methods yield results consistent with previous studies. This ensures an internally consistent parameter scale and maintains continuity across the transition from cool to hot stars within the Ariel sample.}
   {The derived parameters provide an internally consistent basis for studying the link between stellar properties and planetary characteristics in intermediate-mass stars (1.5 < M < 2.32 M$_{\odot}$). Building on our previous work on FGK host stars, we show that correlations between stellar mass, metallicity, and planetary radii also extend to early-type stars, and stellar properties influence the architecture of multi-planet systems.}

   \maketitle
   \nolinenumbers 
%
\section{Introduction}\label{sec:intro}

In the context of planetary science beyond the Solar System, the host stars of exoplanetary systems have become central to our understanding of planets and their environments. Over the past decade, it has grown increasingly clear that the nature of planets, their origins, and their subsequent evolution cannot be fully addressed without considering their host stars. Accurate characterisation of exoplanet host stars is therefore essential for interpreting planetary systems and understanding the conditions under which planets form and evolve. This is particularly true in the field of exoplanetary atmospheres, where a coherent and complete characterisation of the host star is indispensable. 
Planetary parameters -- for instance, the planetary radius, mass, and hence density and bulk elemental composition -- depend on the corresponding stellar values \citep[e.g.][] {Johnson2010,Buchhave2014,Santos2017,Teske2024,DeLaverny2025}. In addition, stellar activity can drive chemical changes in planetary atmospheres or even trigger planetary mass loss, further underscoring the importance of precise stellar characterisation \citep[e.g.][] {Segura2010,Vidotto2015,Johnstone2019,Gupta2023,Nicholls2023}.
To meet these challenges, the community has devoted significant effort to characterising the stars with known planetary systems. However, if the ultimate goal is to trace the formation and evolution of planets, the next step is to extend this effort to a population-wide scale. Such studies require stars and planets to be analysed in a homogeneous way, in order to eliminate systematic errors that arise from, for example, different spectroscopic pipelines or inconsistent photometry. Without this homogeneity, methodological offsets may mimic or obscure genuine astrophysical trends between stellar and planetary properties. Consistency across samples is therefore critical for robust statistical conclusions and for ensuring that observed correlations reflect astrophysics rather than analysis artefacts \citep[e.g.][] {Adib2015,Adib2021,Magrini2022,Biazzo2022,daDilva2024,Filomeno2024}.
Building on this need for a large-scale stellar and planetary characterisation, the first space mission dedicated to a systematic atmospheric population survey will be the ESA M4 mission Ariel. Ariel will observe the atmospheres of roughly one thousand exoplanets, using a tiered survey strategy \citep{Tinetti2018, Tinetti2022}, with the goal of identifying the origins of different planetary classes and ultimately uncovering population-level trends that will advance our understanding of planetary formation and evolution.

To achieve such a significant goal, a precise and homogeneous determination of the stellar parameters is required well before launch in 2029 \citep{Danielski2022}. In fact, to optimise the scientific outcome of Ariel throughout its nominal mission lifetime, a thorough study of the Ariel candidate planetary systems needs to be performed beforehand. This task is essential for identifying key or compelling targets, as well as for increasing the diversity of the population in terms of both planetary atmospheric characteristics and host star properties. By expanding this diversity, we improve the chances of uncovering population-level trends in exoplanet atmospheres, while also placing these planetary atmospheres within a broader Galactic context. In this regard, we refer the reader to \cite{Cowan2025} for an overview of strategies to maximise Ariel’s survey yield, as well as a discussion of the caveats associated with selecting the final target list.
The `Stellar Characterisation' Working Group of the Ariel consortium has focused on the uniform determination of parameters for solar-like FGK stars, which represent the majority of the mission’s candidate sample. \citet[][hereafter M22]{Magrini2022} derived kinematical properties, effective temperatures, surface gravities, metallicities,  microturbulent velocities, masses, and radii for $187$ FGK stars using a combination of spectroscopic, photometric, and astrometric data. More recently, \citet[][hereafter T25]{Tsantaki_2025} extended this analysis to fast-rotating FGK stars, presenting parameters for a total of $353$ stars, which include both slow-rotating and fast-rotating stars,  and additionally providing projected rotational and macroturbulent velocities. Ongoing efforts will further expand this work to cooler M-type stars (Maldonado et al., in prep.).

In this work, we present fundamental parameters for $18$ stars with spectral types from early F to A (6800 < \teff\  < 9560 K). These stars, part of the Ariel Tier 1 Mission Candidate Sample (MCS, \citealt{Edwards2019,EdwardsTinetti2022}), have not yet been characterised by the Ariel Stellar Characterisation Working Group. This is because the methods developed for FGK stars in M22 and T25 are not directly applicable to hotter stars. The rapid rotation in A- and early F-type stars leads to significant line broadening and blending, complicating spectral analysis and reducing the precision of parameter estimation. To maintain continuity across the full sample, we benchmarked our procedure against M22 and T25 in the overlapping temperature regime ($\sim$5000 – 7000\,K), where the spectro-trigonometric approach recovers consistent parameters.

Intermediate-mass stars serve as a testbed for planet formation theories, which offer competing predictions regarding occurrence rates: higher gas-giant frequencies due to massive protoplanetary discs \citep[e.g.][] {Kennedy2008,Johnson2010} versus increased rarity due to rapid disc dispersal or less efficient inward migration \citep{Reffert2015, Johnston2024}. Extending internally consistent characterisation to our sample (\mstar~$ \geq 1.4\, M_\odot$) also allows us to test physical scaling relations, such as the correlation between host-star metallicity and planet presence \citep{Gonzalez1997,Santos_2004,FischerValenti2005} or the dependence of planetary radius inflation on stellar irradiation \citep{Demory2011,Weiss2013}.

\begin{figure}[!htb]
  \centering
  \includegraphics[width=\columnwidth]{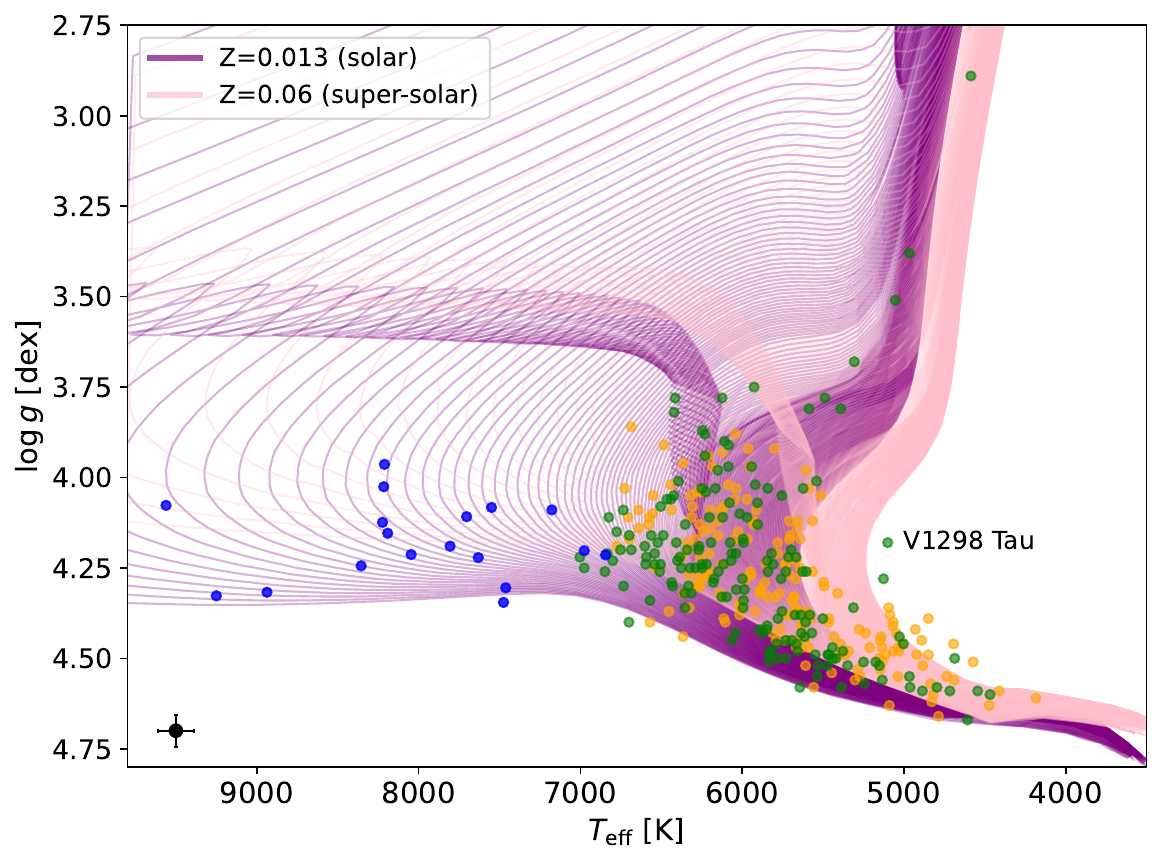}
  \caption{Kiel diagram of analysed stars. Stars analysed in this work are shown as blue dots, those from M22 as orange stars, and those from T25 as green symbols.  The two grids correspond to PARSEC isochrones with ages from 0.1 to 14 Ga, in steps of 0.05 Ga, at solar metallicity (Z = 0.013, in purple) and at super-solar metallicity (Z = 0.06, in pink). The cross in the lower left corner indicates representative $1\sigma$ uncertainties for this work ( $\Delta T_\mathrm{eff}=110\,$K and $\Delta\log g = 0.04\,\mathrm{dex}$).}
  \label{fig:Kiel}
\end{figure}

The sample is described in Sect. \ref{sec:sample}, followed by a description of our spectroscopic method and validation in Sect. \ref{sec:method}. Results are presented in Sect. \ref{sec:results}, with a discussion in Sect. \ref{sec:discussion} and conclusions in  Sect. \ref{sec:summary}.

\section{Sample of stars}\label{sec:sample}
 We selected the 18 hottest stars from the MCS list that have high-quality spectra, except for HATS-70\,\footnote{The FEROS spectrum of HATS-70 exhibited significant noise and spurious spikes, particularly in line cores and above the continuum level. To recover a spectrum suitable for analysis, we applied a custom cleaning procedure combining median filtering, sigma clipping, hard flux capping, and targeted removal of line-core spikes.}. The rest of the hot stars not included in this work are planned to be observed in current or future observing programmes. Figure~\ref{fig:Kiel} shows the Kiel diagram of stars analysed in this work together with stars analysed in M22 and T25.

We selected 23 stars for benchmarking our analysis with M22 and T25. The stars were selected randomly from a temperature range of 4900\,K - 7000\,K and rotational velocities up to $70$\,km\,s$^{-1}$ (Table~\ref{BM_tabel}).

All the spectra were collected by the Ariel Science Consortium Working Group (SCWG) and were accessed via the Ariel Data Drive (Ariel DD). For most of the targets, high signal-to-noise ratio (S/N) spectra were available in public archives (e.g. the ESO Archive Science Portal). A few targets in the hot sample have spectra from successful proposals written by the working group.

We used high-resolution spectra ($R \simeq 48{,}000 - 140{,}000$) collected from the archives and instruments listed in Table~\ref{tab:observations}. As shown by \citet{Tsantaki_2025} and \citet{daDilva2024}, the use of different high-resolution spectrographs introduces only minor variations in the derived stellar parameters. Since our spectra were obtained with the same instruments as in these previous works, we expect any such instrumental effects to be negligible within the quoted uncertainties.

Most spectral regions used in this analysis cover the wavelength range from $\sim 4300$\,\AA\ to $\sim 6500$\,\AA, with combined S/N values ranging from $\approx 100$ to $1700$. Table~\ref{tab:observations} summarises the spectra, including their sources, instruments, wavelength span used, and S/N values.

\begin{table*}[!htb]
\caption{List of the $18$ hot stars analysed in this work.}
\centering
\begin{tabular}{lllllll@{}}
\hline\hline
\# &Star ID     & Spectrograph& S/N     & Analysis $\lambda$ range  (\AA)  & Source (Programme ID)  \\
\hline
1 &HAT-P-57      & HARPS-N& $135$$^c$    & $4397-6429$         & Ariel WG proposal (A41TAC\_45)     \\
2 &HAT-P-70      & ESPRESSO& $617$       & $4397-6429$         & ESO Archive (112.25UT)            \\
3 &HATS-70*      & FEROS& $105$$^c$      & $4397-5867$         & Ariel DD (092.A-9008(A),094.A-9029(G))                \\ 
4 &HD\,2685      & FEROS& $102$$^c$      & $4393-6260$         & ESO Archive (0101.A-9008)         \\
5 &KELT-9        & HARPS-N& $531$$^c$          & $4397-6429$         & Ariel DD (A35DDT4)          \\
6 &KELT-17*      & HARPS-N& $469$$^c$    & $4397-6436$         & Ariel DD (CAT18B\_62)             \\
7 &KELT-20       & HARPS-N& $501$$^c$        & $4397-6429$         & Ariel DD (CAT17A\_38)         \\ 
8 &KELT-21       & HARPS-N& $211$$^c$        & $4478-6443$         & Ariel DD (CAT19A\_97)         \\
9 &KOI-13        & SOPHIE& $594$$^c$     & $4397-6429$         & Ariel DD (11A.PNP.MOUT)           \\
10&MASCARA-1     & HARPS-N& $506$$^c$         & $4397-6429$         & ARIEL DD (A35TAC\_26)        \\
11&MASCARA-4    & ESPRESSO& $1615$      & $4397-6429$         & ESO Archive (0104.C-0605)          \\
12&TOI-615       & HARPS& $269$$^c$      & $4397-6501$         & ESO Archive (108.22LR)            \\
13&TOI-1431*     & HARPS-N& $512$$^c$        & $4397-5869$         & Ariel DD (ITP19\_1)           \\
14&TOI-1518      & HARPS-N& $348$$^c$    & $4397-6442$         & Ariel WG proposal (A46TAC\_27)    \\
15&WASP-33       & HARPS-N& $500$$^c$    & $4397-6443$         & Ariel DD (A34TAC\_42)             \\
16&WASP-172      & ESPRESSO& $204$       & $4397-6449$         & ESO Archive(109.22Z4)             \\ 
17&WASP-178*     & ESPRESSO& $691$       & $4397-6429$         & ESO Archive (108.22GS)            \\
18&WASP-189      & ESPRESSO& $1753$      & $4397-6429$         & ESO Archive (0103.C-0889)         \\
\hline
\end{tabular}
\tablefoot{For each target, we list the spectrograph used for spectral analysis along with the S/N of the spectrum, the wavelength range used for analysis in \AA\,, and the data origin. The spectra were obtained by our collaboration either via successful proposals or the collection of archive data (indicated by programme ID in parenthesis). Stars marked with an asterisk (*) are known as chemically peculiar (e.g. Am) stars. The superscript $^c$ denotes that the shown S/N is derived from combined spectra.}
\label{tab:observations}
\end{table*}

\begin{table*}
\centering
\caption{Fundamental parameters of hot stars derived in this work.}
\begin{tabular}{l c c c c c c c r}
\hline\hline
ID & $T_{\rm eff}$ [K] & $\log g$ [dex] & Mass [$M_\odot$] & Radius [$R_\odot$] & [Fe/H] & [M/H] & $\xi$ [km/s] & $v \sin i$ [km/s] \\

\hline
HAT-P-57   & 7480 $\pm$ 120 & 4.35 $\pm$ 0.06 & $1.49^{+0.03}_{-0.03}$ & $1.40^{+0.05}_{-0.04}$ & -0.20 $\pm$ 0.11 & -0.18 $\pm$ 0.06 & 3.7 $\pm$ 0.3 & 94.3 $\pm$ 1.7 \\
HAT-P-70   & 8220 $\pm$   140 & 4.12 $\pm$ 0.08 & $1.99^{+0.10}_{-0.11}$ & $2.03^{+0.11}_{-0.10}$ &  0.19 $\pm$ 0.27 &  0.15 $\pm$ 0.21 & 3.8 $\pm$ 0.7 & 94.5 $\pm$ 1.7 \\
HATS-70*   & 8190 $\pm$ 180 & 4.15 $\pm$ 0.06 & $1.95^{+0.06}_{-0.07}$ & $1.94^{+0.08}_{-0.07}$ &  0.46 $\pm$ 0.29 &  0.32 $\pm$ 0.12 & 2.9 $\pm$ 1.8 & 42.0 $\pm$ 2.1 \\
HD-2685    & 6840 $\pm$  70 & 4.21 $\pm$ 0.06 & $1.41^{+0.04}_{-0.03}$ & $1.55^{+0.05}_{-0.06}$ &  0.05 $\pm$ 0.15 &  0.03 $\pm$ 0.07 & 1.8 $\pm$ 0.1 & 16.1 $\pm$ 0.4 \\
KELT-17*   & 7630 $\pm$ 120 & 4.22 $\pm$ 0.06 & $1.66^{+0.05}_{-0.06}$ & $1.66^{+0.06}_{-0.05}$ &  0.49 $\pm$ 0.11 &  0.16 $\pm$ 0.11 & 3.3 $\pm$ 0.2 & 46.2 $\pm$ 0.5 \\
KELT-20    & 8940 $\pm$  110 & 4.32 $\pm$ 0.02 & $1.91^{+0.02}_{-0.02}$ & $1.60^{+0.02}_{-0.02}$ & -0.03 $\pm$ 0.09 & -0.02 $\pm$ 0.08 & 2.8 $\pm$ 0.3 & 116.4 $\pm$ 1.2 \\
KELT-21    & 8360 $\pm$  60 & 4.24 $\pm$ 0.05 & $1.96^{+0.04}_{-0.04}$ & $1.77^{+0.05}_{-0.05}$ &  0.11 $\pm$ 0.13 &  0.15 $\pm$ 0.08 & 2.4 $\pm$ 0.4 & 145.9 $\pm$ 2.2 \\
KELT-9     & 9560 $\pm$ 180 & 4.08 $\pm$ 0.03 & $2.32^{+0.04}_{-0.04}$ & $2.33^{+0.04}_{-0.05}$ &  0.09 $\pm$ 0.06 &  0.16 $\pm$ 0.07 & 2.3 $\pm$ 0.2 & 113.9 $\pm$ 1.0 \\
KOI-13     & 8210 $\pm$ 160 & 3.96 $\pm$ 0.06 & $2.02^{+0.05}_{-0.05}$ & $2.46^{+0.10}_{-0.09}$ &  0.21 $\pm$ 0.15 & -0.13 $\pm$ 0.12 & 3.1 $\pm$ 0.4 & 72.1 $\pm$ 0.9 \\
MASCARA-1  & 7700 $\pm$ 120 & 4.11 $\pm$ 0.04 & $1.76^{+0.04}_{-0.03}$ & $1.96^{+0.05}_{-0.05}$ & -0.07 $\pm$ 0.11 & -0.01 $\pm$ 0.08 & 2.9 $\pm$ 0.5 & 105.8 $\pm$ 1.8 \\
MASCARA-4 & 8050 $\pm$ 80 & 4.21 $\pm$ 0.02 & $1.77^{+0.02}_{-0.02}$ & $1.74^{+0.02}_{-0.02}$ &  0.05 $\pm$ 0.09 &  0.05 $\pm$ 0.07 & 2.7 $\pm$ 0.5 & 43.4 $\pm$ 0.5 \\
TOI-1431*  & 7550 $\pm$  90 & 4.08 $\pm$ 0.04 & $1.78^{+0.05}_{-0.06}$ & $2.02^{+0.06}_{-0.05}$ &  0.39 $\pm$ 0.08 &  0.22 $\pm$ 0.13 & 2.9 $\pm$ 0.4 &  7.8 $\pm$ 0.1 \\
TOI-1518   & 7810 $\pm$ 140 & 4.19 $\pm$ 0.05 & $1.85^{+0.03}_{-0.04}$ & $1.82^{+0.06}_{-0.06}$ &  0.26 $\pm$ 0.08 &  0.23 $\pm$ 0.06 & 3.0 $\pm$ 0.2 & 78.4 $\pm$ 1.0 \\
TOI-615    & 6980 $\pm$  30 & 4.20 $\pm$ 0.02 & $1.49^{+0.02}_{-0.01}$ & $1.61^{+0.02}_{-0.02}$ &  0.05 $\pm$ 0.05 &  0.07 $\pm$ 0.04 & 2.0 $\pm$ 0.2 & 17.7 $\pm$ 0.1 \\
WASP-172   & 7180 $\pm$  80 & 4.09 $\pm$ 0.03 & $1.68^{+0.02}_{-0.02}$ & $1.95^{+0.04}_{-0.04}$ & -0.02 $\pm$ 0.09 &  0.09 $\pm$ 0.05 & 2.9 $\pm$ 0.3 & 14.5 $\pm$ 0.3 \\
WASP-178*  & 9250 $\pm$ 130 & 4.33 $\pm$ 0.05 & $2.03^{+0.03}_{-0.03}$ & $1.65^{+0.05}_{-0.04}$ &  0.17 $\pm$ 0.08 &  0.11 $\pm$ 0.08 & 2.6 $\pm$ 0.4 &  9.2 $\pm$ 0.5 \\
WASP-189   & 8220 $\pm$ 120 & 4.03 $\pm$ 0.06 & $2.11^{+0.07}_{-0.07}$ & $2.35^{+0.08}_{-0.08}$ &  0.22 $\pm$ 0.15 &  0.15 $\pm$ 0.12 & 3.3 $\pm$ 0.7 & 95.4 $\pm$ 1.5 \\
WASP-33    & 7460 $\pm$ 80 & 4.30 $\pm$ 0.02 & $1.59^{+0.03}_{-0.02}$ & $1.49^{+0.01}_{-0.02}$ & -0.03 $\pm$ 0.09 & -0.04 $\pm$ 0.09 & 3.8 $\pm$ 0.3 & 84.9 $\pm$ 0.8 \\
\hline

\end{tabular}
\tablefoot{For each star we list the effective temperature (K), surface gravity (dex) with formal ± uncertainties, stellar mass ($M_\odot$), and radius ($R_\odot$) with asymmetric uncertainties, iron abundance [Fe/H] (dex), overall metallicity [M/H] (dex), microturbulent velocity (km/s), and projected rotational velocity (km/s), all with formal ± uncertainties. Stars marked with an asterisk (*) are known as chemically peculiar (e.g. Am) stars.}
\label{tab:Table_hot_results}
\end{table*}

\section{Method}\label{sec:method}

\begin{figure}
\centering
\begin{tikzpicture}[
    node distance=0.2cm,
    every node/.style={font=\small},
    process/.style={
        rectangle,
        rounded corners,
        draw=black,
        align=center,
        minimum width=4.4cm,
        minimum height=0.8cm
    },
    arrow/.style={->, thick}
                        ]

\node (balmer1) [process] {
Balmer line wing fitting\\[-5pt]
\rule{3.8cm}{0.3pt}\\[-2pt]
Initial $T_{\rm eff,B}$ and $\log g$
};

\node (metals1) [process, below=of balmer1] {
Metal-line analysis (fixed $T_{\rm eff,B}$ and $\log g$)\\[-5pt]
\rule{3.8cm}{0.3pt}\\[-2pt]
$[{\rm Fe/H}]$, $[{\rm M/H}]$, $v\sin i$, $v_{\rm mic}$
};

\node (logg1) [process, below=of metals1] {
Trigonometric surface gravity\\[-5pt]
\rule{3.8cm}{0.3pt}\\[-2pt]
$\log g_{\rm trig}$ from $M_\star$, $M_{\rm bol}$, $T_{\rm eff,B}$
};

\node (iter) [process, below=of logg1] {
Metal-line analysis (fixed $\log g_{\rm trig}$)\\[-5pt]
\rule{3.8cm}{0.3pt}\\[-2pt]
$T_{\rm eff}$, $[{\rm Fe/H}]$, $[{\rm M/H}]$, $v\sin i$, $v_{\rm mic}$
};

\node (iter2) [process, below=of iter] {
Final trigonometric surface gravity\\[-5pt]
\rule{3.8cm}{0.3pt}\\[-2pt]
$\log g_{\rm trig}$ from $M_\star$, $M_{\rm bol}$, $T_{\rm eff}$
};

\node (balmer2) [process, below=of iter2] {
Balmer line wing re-fit (fixed $\log g_{\rm trig}$)\\[-5pt]
\rule{3.8cm}{0.3pt}\\[-2pt]
Final $T_{\rm eff,B}$
};

\node (metals2) [process, below=of balmer2] {
Final metal-line analysis (fixed $T_{\rm eff,B}$, $\log g_{\rm trig}$)\\[-5pt]
\rule{3.8cm}{0.3pt}\\[-2pt]
Final $[{\rm Fe/H}]$, $[{\rm M/H}]$, $v\sin i$, $v_{\rm mic}$
};

\draw [arrow] (balmer1) -- (metals1);
\draw [arrow] (metals1) -- (logg1);
\draw [arrow] (logg1) -- (iter);
\draw [arrow] (iter) -- (iter2);
\draw [arrow] (iter2) -- (balmer2);
\draw [arrow] (balmer2) -- (metals2);

\end{tikzpicture}
\caption{Schematic overview of iterative spectro--trigonometric analysis. Boxes represent individual analysis steps, with updated parameters listed below the horizontal line. 
Arrows indicate the workflow from top to bottom. 
Effective temperatures and initial surface gravities were constrained from Balmer line wings, trigonometric surface gravities were derived from stellar mass, bolometric magnitude, and temperature, and metal lines were used to refine all remaining atmospheric parameters until convergence was achieved.}
\label{fig:workflow}
\end{figure}
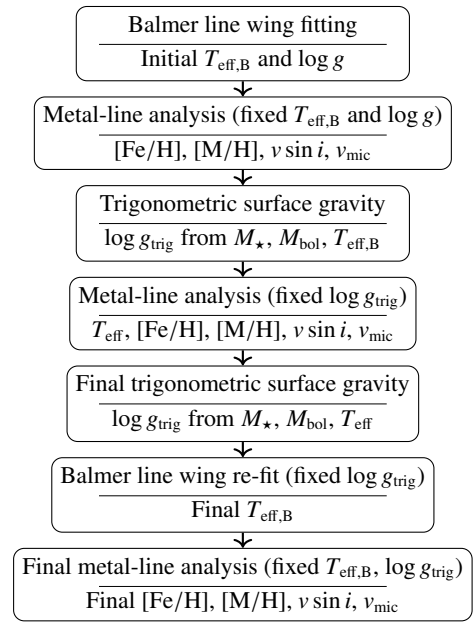

In this work we address the challenges of characterising hot stars. Our methodology follows the iterative spectro-trigonometric approach described in M22 and T25, which combines spectral synthesis, astrometry, and photometry to determine fundamental stellar parameters. We introduced several modifications to the spectroscopic analysis to accommodate the 18 earlier-type stars.
Effective temperatures and surface gravities were initially constrained using Balmer-line fitting, which provides a consistent starting point for the analysis. Metal-line analysis was then performed to refine the remaining atmospheric parameters. Stellar masses were subsequently derived using Bayesian isochrone fitting, and bolometric magnitudes were obtained from {\em Gaia}~Data Release 3 (DR3) parallaxes and broadband photometry. These quantities were combined to compute trigonometric surface gravities. The analysis then iterated: metal-line fitting was repeated with \logg\  fixed to the trigonometric value, and Balmer-line fits were re-done to determine the final effective temperatures, until all parameters converged. This method allowed us to derive effective temperatures (\teff), surface gravities (\logg), metallicities (\meh), iron abundances (\feh), microturbulent velocities (\vmic), projected rotational velocities (\vsini), masses, and radii for $18$ hot stars. A schematic overview of the iterative spectro--trigonometric procedure is shown in Fig.~\ref{fig:workflow}.

\subsection{Comparison to M22 and T25 methods}\label{sec:comp}
M22 employed an equivalent width (EW) method using \textsc{moog}, automated via the \textsc{fama} wrapper \citep{Magrini_2013}. This approach, combined with MARCS model atmospheres, was optimised for FGK-type stars with  $v \sin i$ up to $ \sim 15$\,km\,s$^{-1}$.

T25 extended the parameter space to faster rotators using spectral synthesis with the \textsc{fasma} code\footnote{\url{https://github.com/MariaTsantaki/FASMA-synthesis}} \citep{Tsantaki2018-linelist, Tsantaki_2018}, also based on \textsc{moog}. Their analysis relied primarily on iron lines selected by \citet{Tsantaki2018-linelist}. For Fe lines, the atomic data adopted were those from the {\em Gaia}-ESO  survey \citep{Heiter2021}, as used in M22. For non-iron lines, the atomic data were taken from the \citet{Tsantaki2018-linelist} line list.

In this work, we introduced the following methodological differences: \\
Hot stars exhibit significant line blending and rotational broadening, making fitting isolated iron lines or narrow spectral regions impractical. Instead, we fitted broader wavelength segments starting and ending in continuum regions.

The spectral regions used in this work differ from those in M22 and T25. The regions used are much broader, including many, often blended, spectral lines. Most of the available spectrum between $\sim$4400 \AA\ and H$\alpha$ was used, divided into several large windows that were fit independently.  This approach allows us to use the blended features in the hot star spectra. We excluded wavelength regions strongly affected by telluric contamination, diffuse interstellar bands, and where the continuum was poorly defined (e.g. near Balmer lines).

Regarding metallicity and Fe abundance, M22 and T25 determined Fe abundances and used that as a proxy for bulk metallicity, since their analyses were based on Fe lines.  Our analysis uses lines from a wide range of elements; thus, in addition to iron abundance, we determined a separate \meh, which is based on lines from all elements except for He and Fe. For a chemically normal star \feh\ and \meh\ are expected to be consistent, while any discrepancy may indicate either chemical peculiarities or intrinsic differences in the relative elemental enrichment. The latter naturally occur because not all elements are produced on the same nucleosynthetic timescales or by the same sources in the Galaxy (e.g.\ SNe\,Ia vs\ core-collapse SNe), causing \feh\ and \meh\ to vary across the Galactic population and over time. If a star is chemically peculiar, this choice mitigates the impact of other peculiarities on the Fe abundance.

For the microturbulent velocity, in M22 and T25 it was either derived from spectral analysis or calculated from the \teff, \logg, and \feh\ based on empirical calibrations. We calculated the initial value for \vmic as described in Sect. \ref{sec:method_vmic}, but then fitted it together with the rest of the parameters for all stars.

The Gaia-ESO line list \citep{Heiter2021} used in M22 and T25 was optimised for FGK-type stars and it performed well within that temperature range (up to $\sim$7000 K). However, it was not compiled for hotter stars, and is missing some important lines for them. We extended the \citet{Heiter2021} Gaia-ESO line list by merging it with updated VALD3 \citep{Ryabchikova2015-VALD3} extractions to improve coverage for hotter stars. Empirical oscillator strength corrections were derived for the expanded line list, based on standard stars, described in Appendix\ \ref{Oscillator_strength_corrections}.

For the model atmospheres, for hot stars, due to their broader temperature coverage, we adopted ATLAS9 \citep{Kurucz1993-ATLAS9etc, CastelliKurucz2003}. For benchmark stars, MARCS models \citep{Gustafsson_2008} were used to maintain consistency with M22 and T25. However, that grid of MARCS models only extends to a \teff\ of $8000$\,K.

Finally, spectral synthesis was performed with the \textsc{zeeman} code \citep{Landstreet1988, Wade2001}, which incorporates several recent updates to improve the treatment of Stark and van der Waals broadening, as well as oscillator strength corrections. \textsc{fasma} (used in T25) and \textsc{zeeman} implement similar physical assumptions (LTE, plane-parallel geometry), although they do not use identical partition functions or radiative transfer algorithms. Synthetic spectra from the two codes, for identical input atmospheres and line data, are in excellent agreement as long as Stark broadening is negligible. Importantly, the Fe lines used by T25 are consistently almost identical between the two codes for solar models, and for somewhat higher \teff: small discrepancies become noticeable at 7000 K, which become progressively larger at 8000 and 9000 K.  \textsc{zeeman} includes quadratic Stark broadening using the pre-calculated coefficients in the line list, which are typically more accurate than the general approximation used in \textsc{fasma}. So in hot stars where Stark broadening is important \textsc{zeeman} produces more realistic results. The full analysis procedure and implementation details are described in the following sections.

\subsection{Spectral analysis}\label{sec:Method:spec}
This section builds on the differences outlined above and details the complete analysis procedure applied to the hot star sample.
We derived stellar parameters using the \textsc{zeeman} spectral synthesis code \citep{Landstreet1988, Wade2001,Folsom2012-HAeBe-abun}, which calculates LTE synthetic spectra and fits them to observed spectra using a $\chi^2$ minimisation algorithm \citep{Folsom2012-HAeBe-abun}. 
We used the latest version (20 Apr 2025) of \textsc{zeeman}, which incorporates several improvements, detailed here. 
The Balmer line calculation is described in \citet{Folsom2022}. 
Hyperfine splitting and isotopic splitting are supported through the calculations in VALD \citep{Pakhomov-VALD-HFS}, which can return these components as separate lines.
More advanced van der Waals broadening coefficients, calculated using the theory of \citet{Anstee_OMaral_1995MNRAS.276..859A} and \citet{Barklem_O'Mara_1997MNRAS.290..102B} were used if available.  
Improvements to Stark broadening profiles for some He lines were described by \citet{Folsom2022} \citep[using][]{Barnard1969, Barnard1974, Barnard1975}, which have been supplemented with the Stark broadening tables for He of \citet{Dimitrijevic1990-He-Stark}.  
Continuous opacities were extended to include bound-free and free-free opacity from the H$_{2}^{+}$ and He$_{2}^{+}$ from \citet{Stancil1994} (in addition to existing opacities from H, H$^-$, He {\sc i}, He {\sc ii}, and electron scattering). 
Partition functions approximations from \citet{Irwin1981} were used.

\subsubsection{Line lists and oscillator strength corrections}
\label{sec:lineListCorrections}

As described in Sect. \ref{sec:comp}, we created a merged line list with $\log gf$ corrections suitable for both benchmark and hot stars. A full description of the line list merging and  $\log gf$ corrections is provided in Appendix \ref{Oscillator_strength_corrections}. High-quality lines (flagged as `Y') in the original {\em Gaia}-ESO list were included. Lines with an undecided quality (flagged as `U') were compared with lines from VALD, using observations of standard stars (Sun (G2V), Procyon (F5IV-V), and 21 Peg (B9.5V)) and synthetic spectra, and the data that better reproduced the observations were retained. Lines flagged as not recommended (`N') were rejected. Lines forming their `background line list', (from VALD version 820, in Sept 2014) were replaced by updated entries from VALD. 

The $\log gf$ corrections were applied to all hyperfine or isotropic components of each line, resulting in 2750 corrections for the full list of 26665 lines available on Zenodo\footnote{\url{https://doi.org/10.5281/zenodo.18230682}}.  
These empirical oscillator-strength corrections were applied to approximately 10\% of transitions only when high-quality atomic data were unavailable. No recommended `Y' lines from the {\em Gaia}-ESO or reliable NIST values were modified.

\subsubsection{Normalisation}
Before the fitting procedure, all spectra were continuum-normalised using a custom Python tool that fits low-order polynomials to individual echelle orders \citep{Folsom2025-SpecpolFlow}\footnote{\url{https://github.com/folsomcp/normPlot}}. Since the original echelle order boundaries were not available for most spectra, and because the majority of hot-star spectra exhibit relatively smooth continua, we divided the spectra into uniform $100$\,\AA\ segments for polynomial normalisation. This approach is robust and computationally efficient for smooth continua. For spectra with more complex continuum shapes (e.g. ESPRESSO spectra), the \textsc{suppnet} python code was used \citep{Rozanski2022-SUPPNet}\footnote{\url{https://github.com/RozanskiT/suppnet}}. Balmer lines were normalised using a different approach (see below).

\subsubsection{Balmer line analysis}
Balmer lines provided initial estimates of $T_\mathrm{eff}$ and $\log g$ for all stars, and independent $T_\mathrm{eff}$ values in the final result. For A-type and early F-type stars, Balmer line wings are sensitive to temperature, with only a weak dependence on surface gravity \citep[e.g.][]{Gray2008}, and are relatively insensitive to other parameters such as metallicity, microturbulent velocity, and projected rotational velocity. This makes them valuable as an independent constraint, especially for hot, rapidly rotating stars where metal lines are sparse or blended. In order to minimise the impact of non-LTE effects in the line cores, we removed them. We simultaneously fitted the continuum, $T_\mathrm{eff}$, and $\log g$ to the unnormalised spectrum around each Balmer line, while modelling the local continuum with a second-degree polynomial. As continuum placement and $T_\mathrm{eff}$ are not strictly independent, we tested the approach on standard stars and verified that it yields consistent effective temperatures and reliable initial \logg. We fitted individual lines from H$\alpha$ to H$\delta$, when available, and used their average value. The reported uncertainties are from the standard deviation across the independently fitted Balmer lines for each star. Examples of Balmer line fits are shown in Figure~\ref{fig:KELT21_balmer}.

\subsubsection{Metal-line analysis}
After the initial guess of \teff\ and \logg\ , we obtained other parameters using spectral fitting. Each spectrum was divided into multiple windows for independent analysis. To ensure internal consistency, we kept the windows roughly the same for all hot stars. While the choice of window boundaries does not significantly affect the final atmospheric parameters, it does influence the associated uncertainties. Due to severe line blending in the spectra of hot stars, we used four windows, edges depending on the star, each spanning $\approx$ 300 – 470 \AA. In contrast, the cooler benchmark stars exhibit denser spectral features, so we decided to use narrower windows of approximately $200$ \AA. Within each window, the atmospheric parameters were fitted simultaneously following the scheme described in Sect. \ref{sec:method} and illustrated in Fig.~\ref{fig:workflow}. Depending on the iteration step, $T_\mathrm{eff}$ was either held fixed or allowed to vary together with $[\mathrm{Fe/H}]$, $[\mathrm{M/H}]$, $v\sin i$, and $v_{\rm mic}$. When required, the local continuum was re-fitted using Chebyshev polynomials to account for normalisation imperfections. Initial parameter inputs required for the model atmosphere calculations (namely \teff, \logg, \vsini) were adopted from the Ariel MCS list when available. The initial microturbulent velocity, $v_\mathrm{mic}$, was estimated as described in \ref{sec:method_vmic}, and the solar metallicity was adopted as an initial guess for the metallicity. These values serve only as starting points for the spectral fitting. For all hot stars, Balmer-line fits provided final effective temperatures. We averaged results across all windows to determine final parameters, adopting the standard deviation as the uncertainty to reflect consistency across windows. The final parameters are provided in Table~\ref{tab:Table_hot_results}.

\subsubsection{NLTE effects}
Departures from local thermal equilibrium (LTE) primarily affect \ion{Fe}{i} lines in A--F-type stars through over-ionisation. \citet{Mashonkina_2011} reported non-LTE corrections of $0.02$--$0.10$\,dex for \ion{Fe}{i} and $<0.03$\,dex for \ion{Fe}{ii}, with the magnitude of the effect decreasing towards higher effective temperatures. Since our analysis relies mainly on the wings of the Balmer lines to derive $T_{\mathrm{eff}}$ and on trigonometric $\log g$, such {non-LTE effects are not expected to significantly influence our final stellar parameters.

\subsubsection{Micro-and macroturbulence velocity}\label{sec:method_vmic}

For stars with $T_\mathrm{eff} < 6500$\,K, the initial values of $v_\mathrm{mic}$ were estimated using the empirical relation from \citet{Tsantaki_2013}:
\begin{equation}
v_\mathrm{mic} = 6.932 \times 10^{-4} \cdot T_\mathrm{eff} - 0.348 \cdot \log g - 1.437.
\end{equation}
For hotter stars, initial values were adopted from \citet{Gebran2014}, with typical uncertainties of~$25$\%, and further refined during spectral fitting.  

Macroturbulence velocity ($v_\mathrm{mac}$) was treated differently for hot and cool stars. For benchmark stars, we adopted empirical relations from \citet{Doyle2014}, consistent with T25. These are valid for dwarfs with $5200\,\mathrm{K} < T_\mathrm{eff} < 6400\,\mathrm{K}$ and $\log g > 4.0$. Similar to T25, for benchmark stars falling out of these ranges, we followed \citet{ValentiFicher_vmac}.
  
For the hot star sample, the macroturbulence velocity was set to zero, as rotational broadening dominates line profiles in this regime. Although some studies (e.g. \citealt{Doyle2013_vmacWASP}; \citealt{Psaridi2023}) extrapolate macroturbulence from \citet{Gray2008} up to $\sim$6700\,K, these relations remain uncertain, particularly for A-type stars. We note that setting $v_\mathrm{mac} = 0$ may lead to a mild overestimation of $v \sin i$, especially with host stars with low \vsini\ such as WASP-178, but the value remains within error margins and does not affect the determination of other stellar parameters.

\subsubsection{Trigonometric surface gravity}

Similarly to M22 and T25, surface gravity was obtained using the following relation:
\begin{equation}
\log g_\mathrm{trig} = \log M_{\star} + 0.4 \cdot M_\mathrm{bol} + \log T_\mathrm{eff} - 12.505,
\label{eqn:logg}
\end{equation}
where $M_{\star}$ is the stellar mass in solar mass units, $M_\mathrm{bol}$ is the bolometric magnitude, $T_\mathrm{eff}$ is the effective temperature adopted at each iteration (initially from Balmer-line fitting and updated during the iterative analysis), and $12.505$ is a normalisation factor to the solar surface gravity.
The bolometric magnitude, $M_\mathrm{bol}$ (and consequently the luminosity), was derived by a Markov chain Monte Carlo (MCMC) -based fitting procedure, which takes input as a combination of several magnitudes and the distance. This procedure does not perform a spectral energy distribution (SED) fit; instead, the MCMC sampler predicts theoretical magnitudes from bolometric corrections and compares them to the observed photometry.
The magnitudes considered are taken from 2MASS ($J$, $H$, and $K_s$ bands; \citealt{Skrutskie06}), {\em Gaia} ($G_\mathrm{BP}$ and $G_\mathrm{RP}$; \citealt{GAIAEDR3}), and, when available, Johnson $B$ and $V$ magnitudes. The Johnson magnitudes are taken from the synthetic photometry generated from {\em Gaia} BP/RP mean spectra \citep{Montegriffo+23}.
Each magnitude was converted to bolometric magnitude using YBC bolometric correction library \citep{YBC}, which interpolates a series of precomputed tables of bolometric correction in \teff, [Fe/H], \logg\, and extinction, $A_V$. In this specific step, $\log g_{\rm trig}$ was used only to determine the bolometric corrections and was not treated as an evolutionary constraint. The distance was derived by a simple parallax inversion, with the parallax given by {\em Gaia} {\sc EDR3}.
The masses were derived using the Bayesian tool {\sc PARAM} \citep{PARAM,PARAM2,PARAM3}, which performs isochrone fitting by comparing observed parameters (such as metallicity, effective temperature, and photometric luminosity) with a grid of theoretical stellar models \citep{Moedas2022}.
Importantly, $\log g_{\rm trig}$ was not fitted to the evolutionary tracks, but was derived independently from the resulting mass, luminosity, and temperature using Equation \eqref{eqn:logg}.
The metal-line fitting was repeated with \logg\ fixed to the trigonometric value, iterating until the parameters converged. All fits were inspected visually. In Figure \ref{fig:obs-synt-diff} we show the observed, model, and difference spectra for the hot stars.

\subsection{Peculiar stars}\label{Pecliar stars}
Several stars in our sample are known chemically peculiar stars: HATS-70, KELT-17, TOI-1431, and WASP-178. Literature reports for these targets indicate either strong or weak Am-type abundances (underabundant Ca and Sc and overabundant iron-peak elements). For these stars, our standard fitting procedure was applied to determine the overall metallicity [M/H] and iron abundance [Fe/H]. However, given the peculiar abundance patterns in such stars, the derived parameters should be interpreted with caution. Am-type chemical peculiarity is generally associated with slow rotation in stars hotter than approximately 7000 K.  In our analysis, we notice a >~0.1 dex difference between \feh\ and \meh\ for these known peculiar stars (see Table \ref{tab:Table_hot_results}). We adopted this offset as a preliminary diagnostic flag for potential peculiarity in other targets. Specifically, stars with lower \vsini\ such as HD-2685, TOI-615, WASP-172, and potentially KOI-13 and TOI-1518 fall within the regime where diffusion-driven abundance anomalies may occur. In addition, WASP-189 was classified as non-Am by \citet{Saffe_2021} and \citet{Anderson2018}, whereas \citet{Lendl2020} reported Am-type peculiarities for this star. A definitive Am classification requires a full spectral classification analysis (e.g. using MK classification methods \footnote{\url{https://www.appstate.edu/~grayro/mkclass/}}) or detailed element-by-element abundance studies, both of which are beyond the scope of this work. These analyses will be performed in future work and may lead to further refinements of fundamental parameters such as $T_\mathrm{eff}$ and possibly $\log g$.

\subsection{Consistency with previous works}\label{sec:Comparison_BM}
\begin{figure}
    \centering
    \includegraphics[width=1\columnwidth]{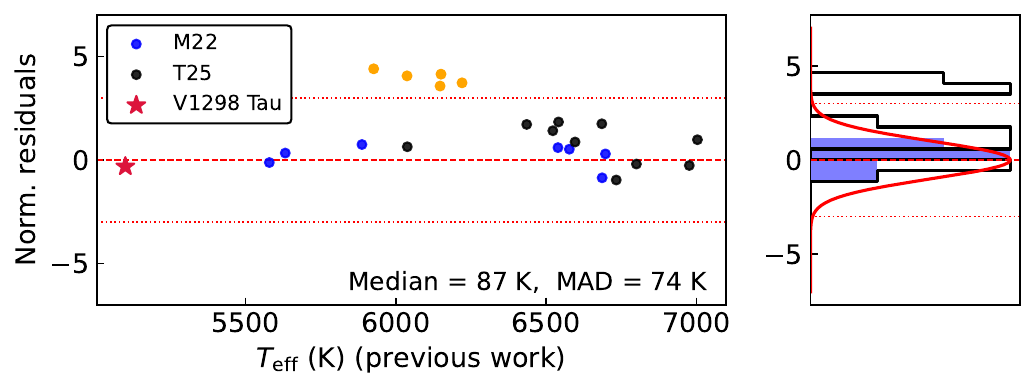} \\
    \includegraphics[width=1\columnwidth]{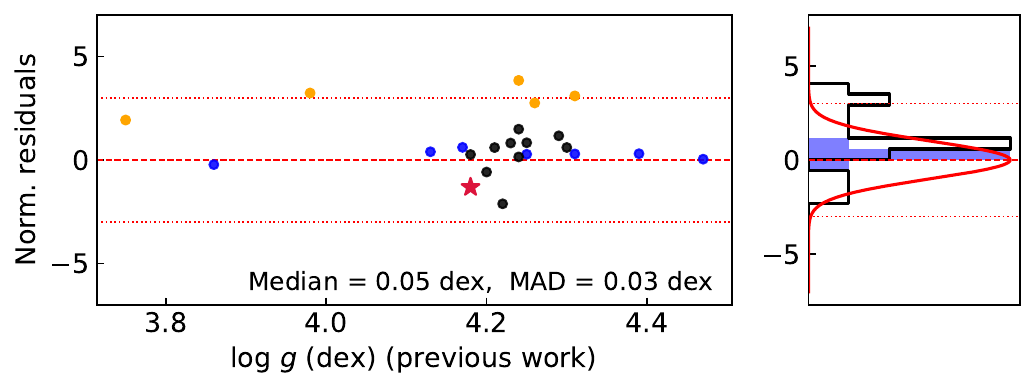}\\
    \includegraphics[width=1\columnwidth]{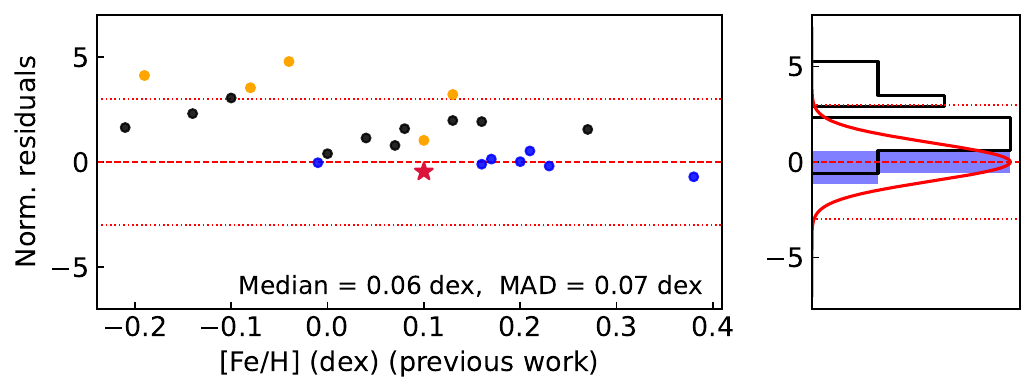}\\
    \includegraphics[width=1\columnwidth]{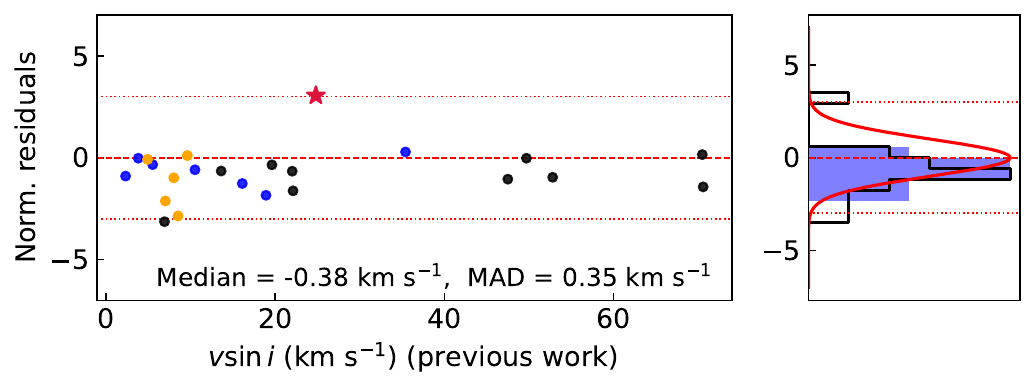}\\
    \includegraphics[width=1\columnwidth]{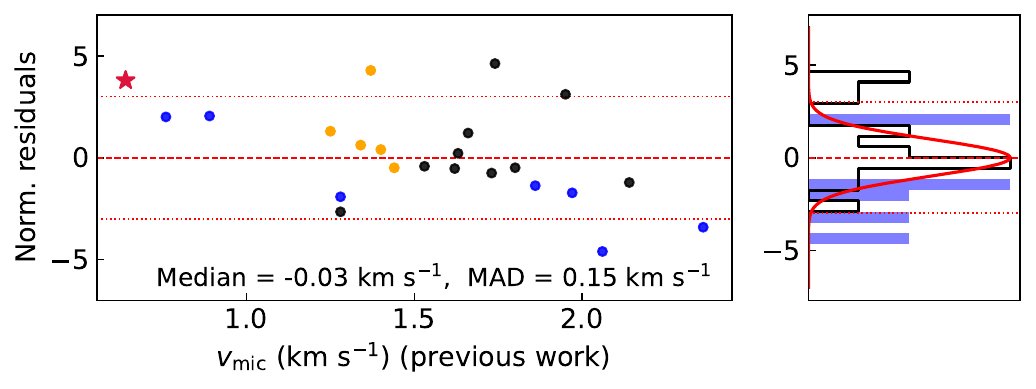}\\
 \caption{\textit{Left panels}: Normalised residuals defined as the difference between this work's estimates and previous ones, divided by the propagated uncertainty of both estimates plotted against previous estimates from M22 and T25. Outliers are defined as stars with normalised residuals lying outside the 3$\sigma$ region, marked by the dashed red lines. Orange points indicate the 5 stars from T25 for which \teff\ is above 3$\sigma$. Median and median absolute deviation (MAD) values are indicated in each panel.
\textit{Right panels: }Distribution of normalised residuals. The red curve shows a standard normal distribution (mean zero, variance one) to ease visual assessment of consistency between works.
}
    \label{fig:Comparison}
\end{figure}

To assess the consistency of our methodology, we analysed a set of $23$ benchmark stars, previously studied in M22 and T25, using the methodology described in Sect.~\ref{sec:Method:spec}. Figure~\ref{fig:Comparison} compares our inferred stellar parameters (\teff, \logg, \feh, \vsini, and \vmic) with those from M22 and T25 (hereafter `previous work'), showing normalised residuals relative to the previous estimates (left panels) and their distributions (right panels). We present the residuals with respect to M22 and T25 separately, as their methodologies, though equivalent, are not identical (see T25 for details). For the benchmark stars, we derived effective temperatures using both Balmer-line and metal-line diagnostics to assess which method is more reliable in different temperature regimes. A comparison with M22 and T25 reveals an empirical transition at \teff$\simeq6600$\,K. Below this temperature, Balmer-line fits systematically overestimate effective temperatures due to increased metal-line blending in the wings. However, above this threshold, Balmer-line and metal-line diagnostics converge, showing excellent agreement (e.g. $0$~K difference for HAT-P-2). This transition point guides our methodology: we adopted Balmer-line temperatures for the hot-star sample to ensure an internally consistent scale that maintains continuity with the cooler stars analysed in previous works. Table \ref{BM_tabel} provides the effective temperatures derived from metal-line analysis ($T_{\rm eff}$) and from Balmer-line fits ($T_{\rm eff,Balmer}$) for all benchmark stars, allowing for a direct comparison between the two diagnostics.

Outliers exist in  \teff, \logg\ , or \feh\ panels, with normalised residuals lying outside the 3$\sigma$ level. The largest scatter among the remaining parameters is seen for \vmic, which displays several points that are clearly inconsistent at the $>3\sigma$ level, suggesting that its uncertainties may be underestimated. 
This increased scatter in $v_{\rm mic}$ is expected in cross-study comparisons. Microturbulent velocity is not a directly observable quantity and is treated differently in previous works,  being either fitted, constrained, or computed from empirical relations (see Sect. \ref{sec:method_vmic}).
In contrast, \vsini\ values appear fairly consistent between the two analyses, although with a small median offset.
The uncertainties in \vsini\ naturally increase with the rotation rate, illustrating the challenges of analysing rapid rotators. \textsc{fasma} performs well for stars up to \vsini\ $\sim 50$\,km/s, while \textsc{zeeman} is used to analyse stars with even higher projected rotational velocities. 
The stars flagged as outliers are HR 858, KELT-2A, TOI-677, K2-99, and WASP-166 (orange points in Fig. \ref{fig:Comparison}), whose normalised residuals in \teff\ exceed 3$\sigma$. These stars also show systematically higher inferred values of \logg\ and \feh, and all five stars seem to have underestimated uncertainties. For \teff, our average uncertainties are 36 K (38 K in T25). For \logg\ and \feh\ 0.03 (0.01 dex T25) dex and 0.04 (0.03 dex T25) dex, respectively.
We also assessed how residuals in \teff, \logg, and \feh\ vary as a function of the M22 and T25 values (Fig.~\ref{fig:Correl_BM_param}). No significant correlations are found, although a weak tendency for higher \teff\ values at lower metallicities is visible (r = – 0.40, p = 0.06).
T25 similarly reported a $\sim145$\,K $T_{\mathrm{eff}}$ overestimation compared to M22 for stars with \feh\  $< - 0.25$\,dex, without significant offsets in \logg\ or \feh. These results suggest that small residual trends may originate from the interplay between $T_{\mathrm{eff}}$ and \feh, while \logg\ remains largely unaffected.

The coolest target in our benchmark set, V1298 Tau, is a pre-main-sequence star published in M25 (marked in Fig.~\ref{fig:Kiel} and also in Fig.~\ref{fig:Comparison} in red). However, given that the isochrone grids used for stellar mass estimation were not calibrated for young stars, it has now been excluded from the Ariel Stellar Catalogue.

\section{Results}\label{sec:results}
Here we present our results for the sample of hot stars in the MCS.
The final sample comprises $18$ stars with spectral types from late F to early A. Consistency checks on the benchmark sample (Sect.~\ref{sec:Comparison_BM}) show that our methodology yields results broadly consistent with those in T25 and M22. Small offsets and trends in \teff\ and \feh\ were mitigated by adopting \teff\ values from Balmer‐line fitting for the final metallicity analysis. With these offsets characterised, our results are placed on the same scale as the cooler‐star analyses in M22 and T25, enabling a consistent comparison of stellar parameters across a wide temperature range for the Ariel MCS.

\subsection{Overview of hot star parameters}
The Kiel diagram (Fig.~\ref{fig:Kiel}) shows the hot star sample alongside the cooler stars from previous works, overlaid with \textsc{parsec} isochrones\footnote{\url{https://stev.oapd.inaf.it/cgi-bin/cmd}} of solar and super-solar metallicity. The new targets occupy the upper main sequence above $\sim6900$\,K. All parameters and uncertainties are listed in Table~\ref{tab:Table_hot_results}. Combined with the cooler FGK stars analysed in M22 and T25, the total number of Ariel MCS stars with internally consistent parameter determinations now stands at $370$.

Figure~\ref{fig:histograms} shows the distributions of \teff, \logg, \feh, and stellar mass and the projected rotational velocities for the hot star sample (shown in red) compared with the M22 and T25 sample (shown in blue). Our sample is concentrated at M$_\star > 1.4$\,M$_\odot$ and extends up to $\sim2.3$\,M$_\odot$, substantially expanding the high-mass tail of the Ariel MCS. The metallicity range is similar to that of the cooler star sample, dominated by solar and supersolar values.

Figure~\ref{fig:vmic} shows the microturbulent velocity as a function of  \teff. The trend follows well-known temperature dependence of microturbulence in A–F stars: \vmic\ rises from late-F to early-A types, reaching a maximum around $T_{\rm eff}\sim8000$\,K, and then decreases towards hotter stars (e.g. \citealt{Landstreet2009}).
Figure~\ref{fig:vsini} shows $v \sin i$ as a function of stellar mass, colour-coded by $T_{\mathrm{eff}}$. The relation between $v \sin i$ and stellar mass follows the expected trend: the rotation rates are higher at larger masses. The fastest rotators are KELT-21, KELT-20, KELT-9, and MASCARA-1, which span masses from  $\sim1.6$ to $2.3$\,M$_\odot$ and effective temperatures from  $\sim8000$ to $\sim9400$\,K. In contrast, TOI-1431 is among the slowest rotators in the sample with $v \sin i \sim 8$\,kms$^{-1}$. The hottest stars in the sample ($T_{\mathrm{eff}}\gtrsim8000$\,K) tend to rotate the fastest, consistent with their thinner convective envelopes and weaker magnetic braking. At a given mass, however, there is significant scatter, with some intermediate-mass stars showing relatively slow rotation ($v \sin i \lesssim 40$\,kms$^{-1}$). Most of these slow rotators also exhibit chemical peculiarities, as mentioned in section \ref{Pecliar stars}.

\begin{figure}
    \centering
    \includegraphics[width=\linewidth]{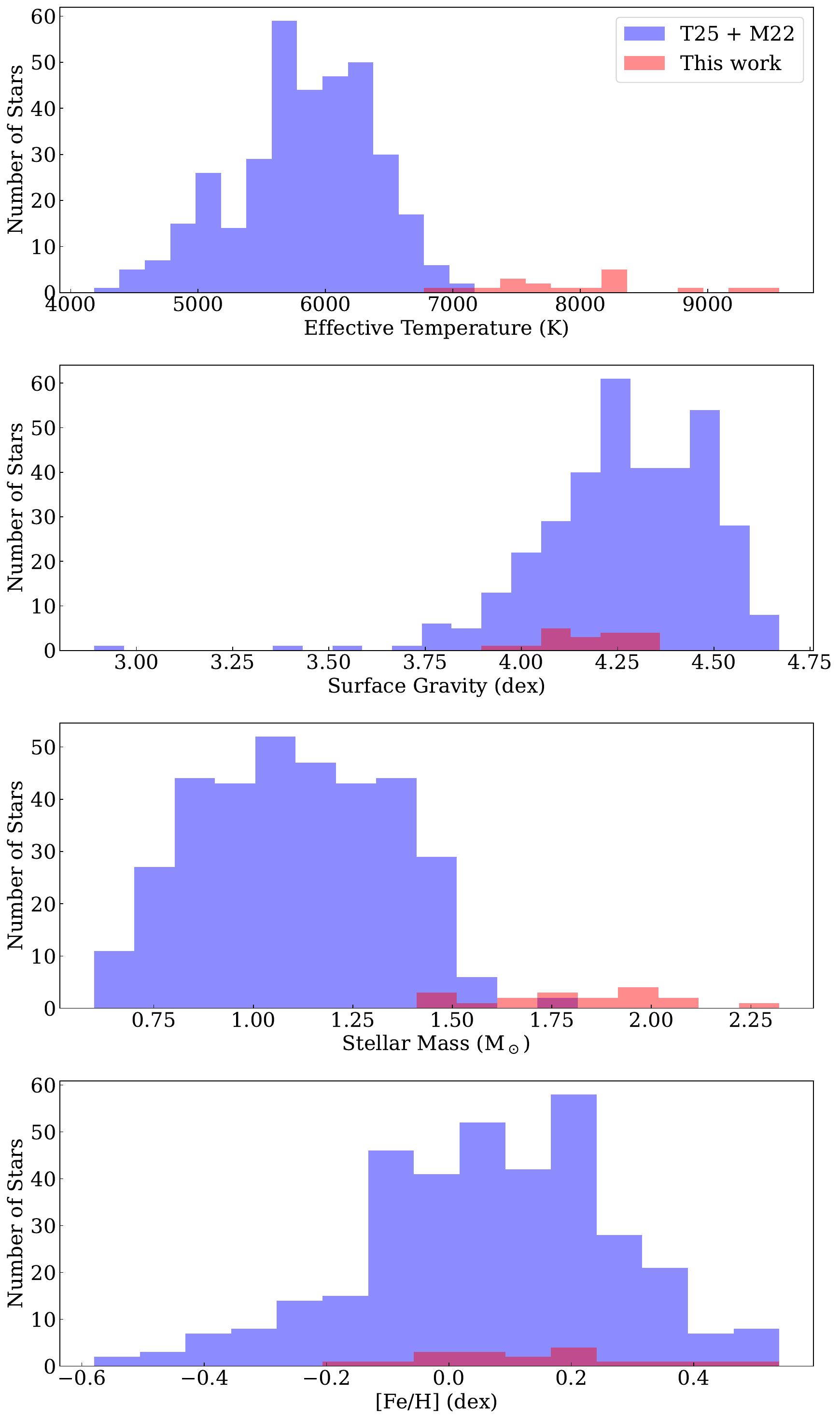}
    \caption{Distributions of homogeneously derived stellar parameters for the Ariel mission candidate sample. The red histograms represent the stars analysed in this work, while the filled blue bins show results from our previous analysis (T25 and M22).}
    \label{fig:histograms}
\end{figure}

\begin{figure}
    \centering
    \includegraphics[width=1\linewidth]{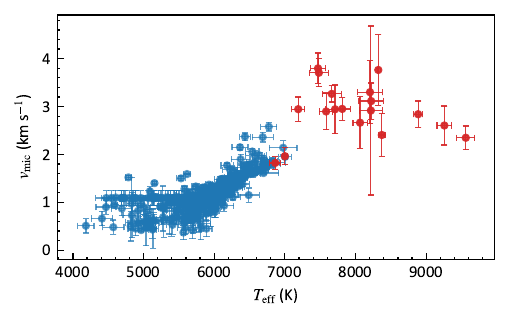}
    \caption{Microturbulent velocity as a function of \teff. Red points are stars from this analysis and blue points are stars from M22 and T25.}
    \label{fig:vmic}
\end{figure}

\begin{figure}[!htb]
    \centering    \includegraphics[width=1\linewidth]{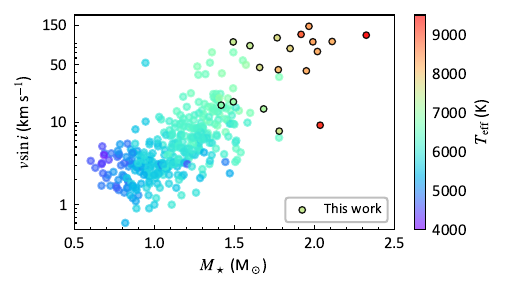}
    \caption{Projected rotational velocity ($v \sin i$) as a function of stellar mass, colour-coded by effective temperature. Hot fast rotators analysed in this work are highlighted with a black circle.} 
    \label{fig:vsini}
\end{figure}

\subsection{Kinematic properties}\label{kin}

Figure~\ref{fig:Toomre} shows the kinematic properties of the hot star sample compared to the T25 sample in the Toomre diagram (left) and the Galactic orbital parameter space (right). All hot stars lie within the thin disc region ($|V_{\rm tot}| < 50$\,\kms), consistent with a young, local population. Their orbital parameters indicate they are located near the solar Galactocentric radius ($R_{\rm GC} \sim 8$\,kpc) with minimal radial excursions, suggesting they have not migrated significantly from their birthplaces. This is expected for massive, short-lived main-sequence stars, which do not survive long enough to experience substantial radial migration.

\subsection{Comparison of hot star parameters with literature values}
\begin{figure}[!htb]
    \centering
    \includegraphics[width=1\columnwidth]{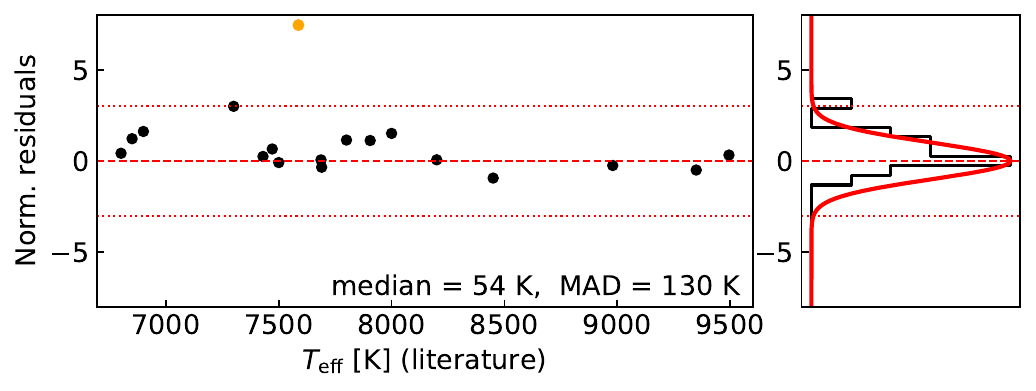} \\
    \includegraphics[width=1\columnwidth]{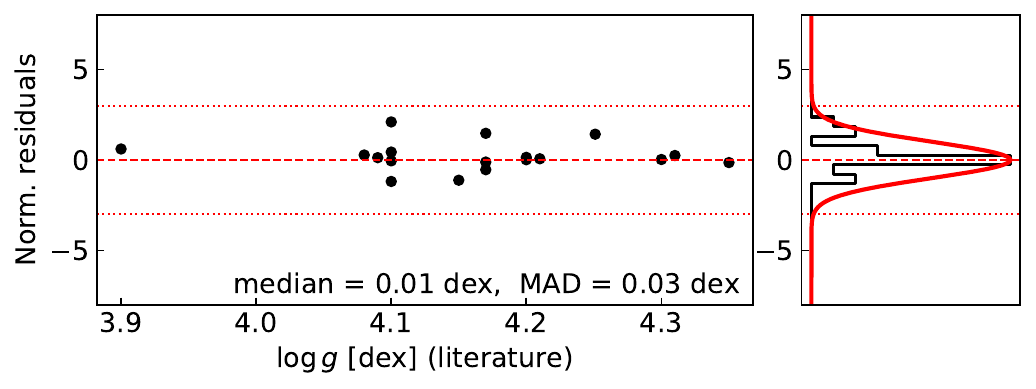}\\
    \includegraphics[width=1\columnwidth]{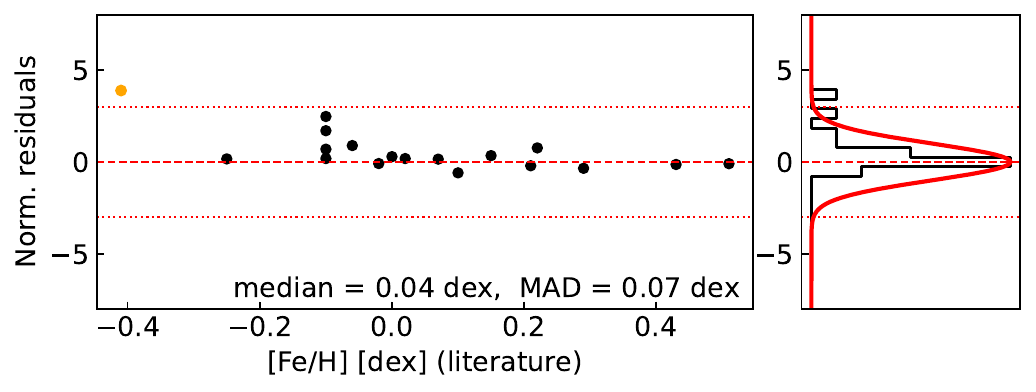}\\
    \includegraphics[width=1\columnwidth]{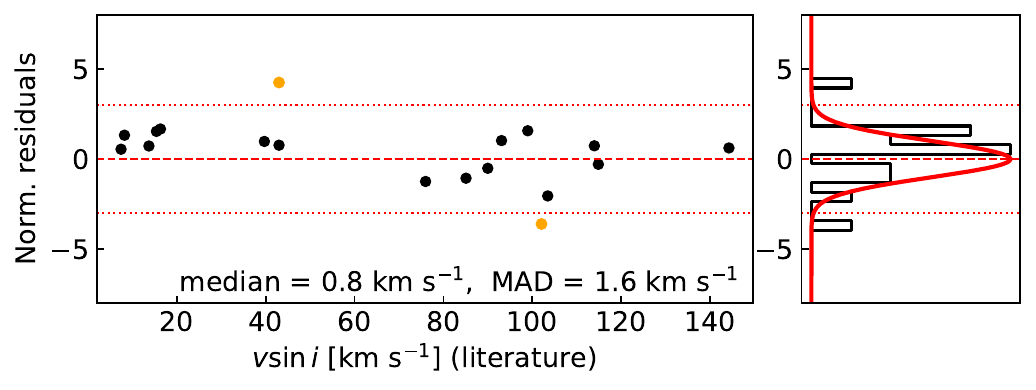}\\
 \caption{\textit{Left panels}: Same as in Fig.\ref{fig:Comparison}, but for 18 hot stars. Literature values are from \cite{Hartman2015,Zhou_2019_HAT-P-70,Jones2019, Niemczura2017,Johnson2018, Kama2023,Addison2021,Cabot2021,Hellier2019,Hellier2019b,Lendl2020,Psaridi2023, Saffe_2021}, and \cite{Saffe_2022}.
\textit{Right panels: }Distribution of normalised residuals. The red curve shows a standard normal distribution (mean zero, variance one).
}
    \label{fig:Comparison_Hot}
\end{figure}

We compared our derived stellar parameters with those available in the literature. In particular, we focused on the works of \citet{Saffe_2021,Saffe_2022}, who have conducted detailed analyses of early-type stars, including both atmospheric parameters and chemical abundances. Similar to us, they employed high-resolution spectroscopy and synthetic spectrum fitting techniques to derive atmospheric parameters. 

We validated our results against literature values using normalised residuals (Fig.~\ref{fig:Comparison_Hot}). Each panel shows the residuals divided by propagated uncertainties as a function of the literature value, with the median and MAD annotated. The majority of stars cluster around zero residuals, confirming consistency with previous works within the quoted uncertainties. A few
$>3\sigma$ outliers are seen: KELT-21 in \teff\  and \feh, and HAT-P-57 and KELT-17 in \vsini.

The literature values for KELT-21 are taken from \citet{Johnson2018}, who used the Payne algorithm (see references therein) to model individual TRES spectra by fitting a $\sim200$\,\AA\ region around the \ion{Mg}{I} triplet near $5200$\,\AA. 
Since we used Balmer line analysis as an independent method of determining \teff\ together with independent \logg\ estimation, our values are expected to be more robust. Our \teff\ is closer to the value \cite{Johnson2018} derived using SED and in better agreement with the \textit{Gaia}~DR3 parameters.  Specifically, \textit{Gaia}~DR3 reports \teff = $8743^{+53}_{-61}\,\mathrm{K}$ and \feh$ = +0.02^{+0.05}_{-0.01}\,\mathrm{dex}$, 
whereas \citet{Johnson2018} derived \teff$ = 7598^{+81}_{-84}\,\mathrm{K}$ and \feh $= -0.405^{+0.032}_{-0.033}\,\mathrm{dex}$. Such a low \feh\ is unlikely for an early A-type thin-disc star and may result from limitations of automated approaches such as the Payne. The results can be affected by high rotational broadening, which reduces sensitivity to metallicity diagnostics and biases fits towards lower values.
As for HAT-P-57 ($94 \pm 1.7$\ \kms\  versus $102 \pm 1.3$ in \cite{Hartman2015} and KELT-17 ($46.2 \pm 0.5$\ \kms\  versus $43.0 \pm 0.6$ in \cite{Saffe_2021}, the outliers in \vsini\ arise from underestimated formal uncertainties rather than real discrepancies.

\section{Discussion}\label{sec:discussion}
Our results extend the known parameter space of stellar hosts towards the high-temperature range, providing the first consistent analysis of A‐type planet hosts within the Ariel sample. Hot stars are particularly relevant for exoplanet studies, as their relatively line‐poor spectra make them favourable targets for transmission and emission spectroscopy. Moreover, giant planets around A‐type stars are intrinsically rare \citep{Borgniet2019}, meaning each system provides a valuable probe of planet formation and migration around intermediate‐mass stars. This new subsample therefore fills an important gap in the Ariel stellar characterisation, linking the hotter end of the main sequence to the cooler FGK hosts analysed in M22 and T25.

In this section, we extend the study of planet–star connections presented in our previous works. We use planetary parameters from the NASA Exoplanet Archive\footnote{\url{https://exoplanetarchive.ipac.caltech.edu}}. The characteristics of the planet and hot host star systems can be found in Table \ref{tab:hj-systems}.

\subsection{Stellar metallicity and exoplanet populations}

\begin{figure}[!htb]
    \centering
    \includegraphics[width=1\linewidth]{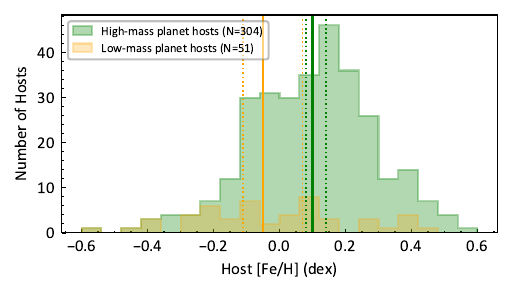}
    \caption{Host star metallicity distributions for Ariel target candidate sample systems with low-mass planets ($M_\mathrm{p}<0.2\,M_\mathrm{J}$; yellow) and high-mass planets ($M_\mathrm{p}\geq0.2\,M_\mathrm{J}$; green). Solid vertical lines mark the median \feh\ values, and the dotted lines indicate the $95\%$ confidence intervals.}
    \label{fig:hist_metal_distribution}
\end{figure}

T25 confirmed that stellar metallicity influences planet populations, using a mass threshold of $M_\mathrm{p} = 0.2 M_\mathrm{J}$ to distinguish between low- and high-mass systems. Their results showed that low-mass planets preferentially orbit metal-poor stars, while high-mass planets typically orbit metal-rich hosts.

All 18 hot stars analysed in this work host hot Jupiters, and therefore fall into the massive-planet category. The inclusion of this new subsample keeps the overall median of the metallicity distribution around $0.1$\,dex (Fig.~\ref{fig:hist_metal_distribution}).  

Furthermore, in Sect.~\ref{kin} we showed that all of our hot stars belong to the thin disc. This is expected, as intermediate-mass A-F stars are relatively young. Their youth and Galactic disc inheritance makes them enriched in heavy elements, providing more building blocks for giant planet formation. For a broader discussion of the implications for thin- and thick-disc stars, we refer to T25. Future detailed abundance analyses of our hot sample will provide a more robust test of Galactic membership.
 
\begin{figure}[!htb]
    \centering
    \includegraphics[width=1\linewidth]{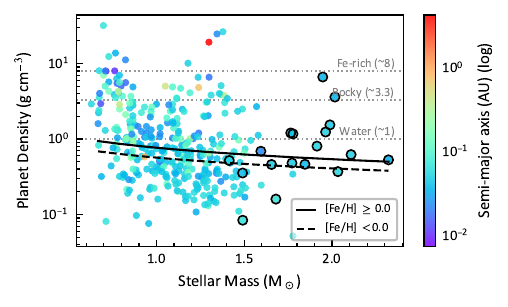}
    \caption{Planet density as a function of stellar mass. The points are colour-coded by semi-major axis (AU), with a logarithmic colour scale. Solid and dashed black lines show linear fits for high- and low-metallicity hosts (\feh\ $\geq$ 0 and < 0, respectively), considering only planets more massive than 0.2 $M_\mathrm{J}$. Following the compositional density divisions from \cite{Zeng2019}, we indicate reference lines at $\rho\sim1, 3.3$, and $8$ g cm$^{-3}$ corresponding respectively to a water- or ice-rich, silicate-rocky, and iron-rich bulk composition. Black circles highlight the 18 stars analysed in this work.}
    \label{fig:pl_density}
\end{figure}

\subsection{Dependence of planetary structure on stellar mass and metallicity}
To capture the combined effects of stellar mass and metallicity, we examined
the relation between planetary density and stellar mass
(Fig.~\ref{fig:pl_density}). Planet density decreases towards higher stellar masses, consistent with the enhanced radius inflation expected for planets orbiting more luminous A–F-type hosts. We still observe a significant scatter among planets around intermediate-mass stars, which is partly explained by uncertainties in planetary masses; six planets in our hot sample have only upper limits on their masses, which artificially raises their inferred densities and flattens the overall planet-density–stellar-mass trend.

At a given stellar mass, planets around metal-rich stars are systematically denser, indicating a larger heavy-element fraction or more massive cores, in agreement with the metallicity–density trend reported by \citet{Biazzo2022}. Together, these results support a picture in which metallicity regulates the internal heavy-element content, while stellar mass primarily drives envelope expansion through irradiation.

To isolate the physical mechanism behind this trend, we examined planetary density as a function of intercepted stellar power
(Fig.~\ref{fig:pl_density_Pinc}). A strong anti-correlation is observed for giant planets: systems receiving higher incident power host less dense planets, demonstrating that irradiation is the dominant driver of radius inflation. Inflation becomes significant once the incident flux exceeds approximately 
\(2 \times 10^{8}\,\mathrm{erg\,s^{-1}\,cm^{-2}}\) (or equivalently 
\(\sim 2 \times 10^{5}\,\mathrm{W\,m^{-2}}\)), the empirical threshold at which
hot-Jupiter radii begin to rise sharply 
\citep[e.g.][]{Demory2011}. Most planets in our hot-star sample
lie well above this threshold, consistent with their large radii.  

At fixed incident power, giant planets around metal-rich stars remain denser, reinforcing the role of metallicity in setting bulk composition. These results extend the trends previously identified for FGK hosts in T25 to the higher-mass A–F stars analysed in this work.
Following T25, we examined radius–stellar mass correlation separately (Appendix~\ref{fig:r_vs_m_2bins}). This is consistent with a scenario where both stellar luminosity and mass both drive planetary radius inflation.

\subsection{Multi-planet systems and their relation with the stellar mass, metallicity, and total planetary mass}  

\begin{figure}[t]
    \centering
     \includegraphics[width=0.95\linewidth]{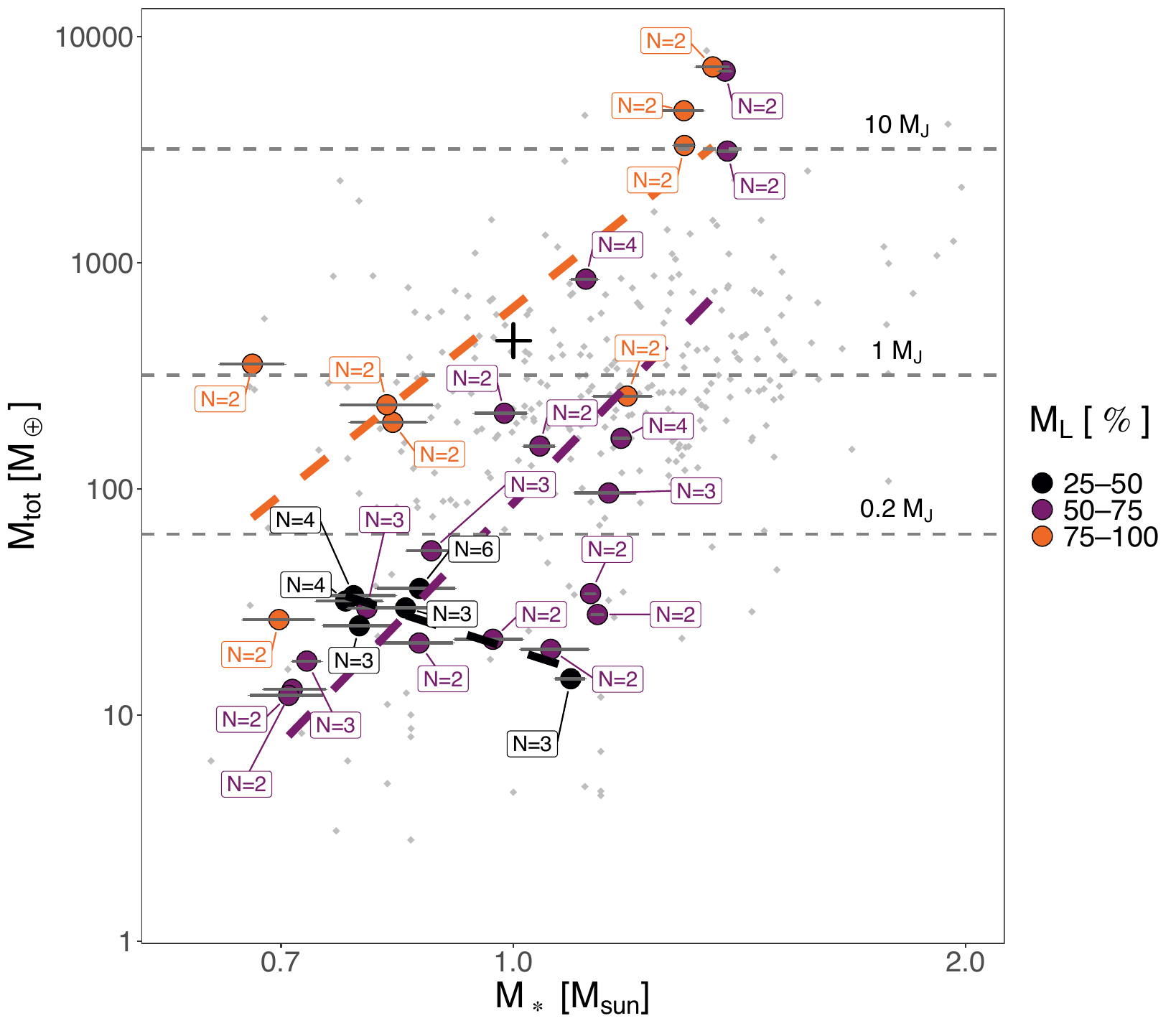}
     \caption{Total planetary mass available in each planetary system, as a function of the stellar mass (with error bars), colour-coded by the $M_L$ metric (see main text). The grey dots represents the mass of planets in single planet systems. Dashed lines are plotted to provide a visual range of masses in terms of \Mj. The black cross identifies the Solar System and stellar mass errors are reported for each system.}
    \label{fig:mtot_mlarge}
\end{figure} 

\begin{figure}[h]
    \centering
   \includegraphics[trim=0.1cm 0cm 2.2cm 0.cm, clip,width=1.02\linewidth]{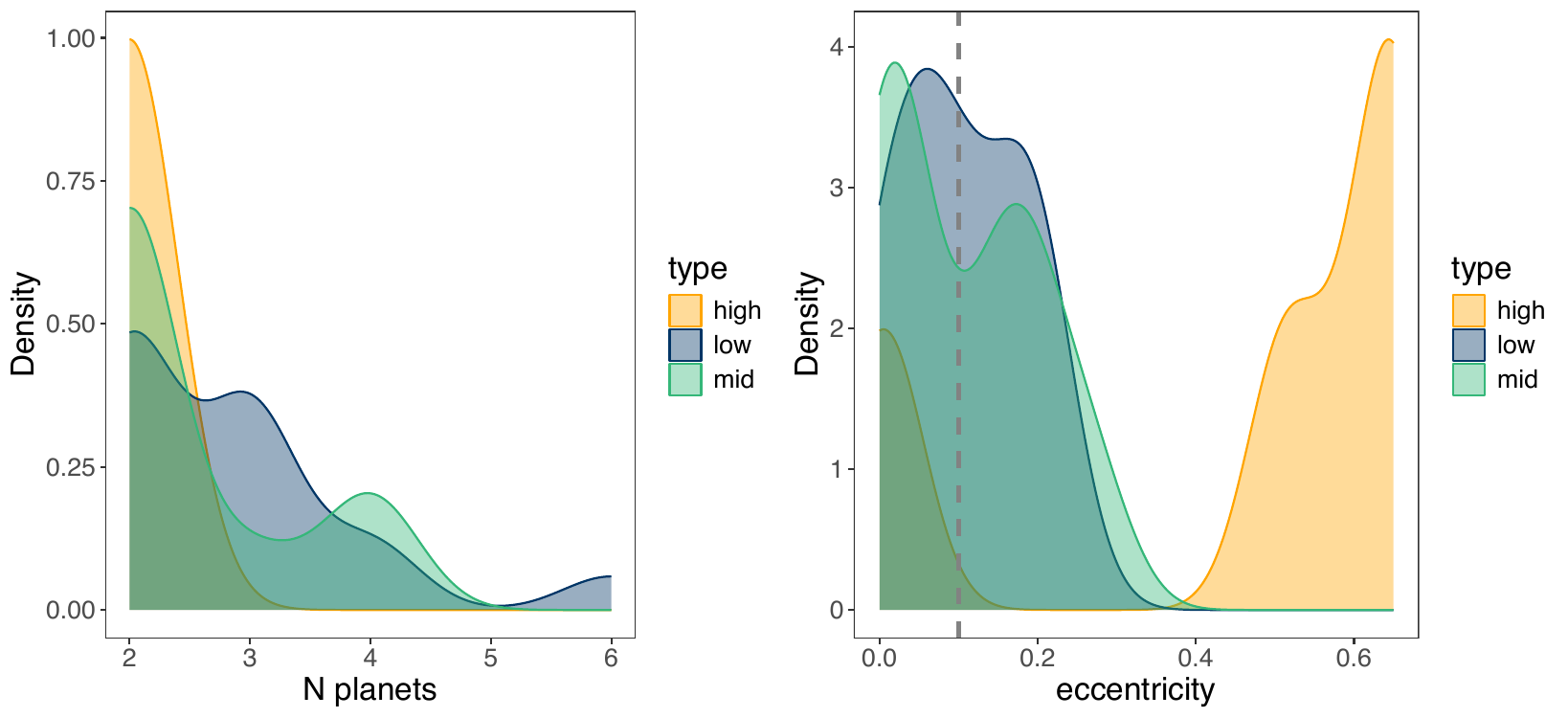}
    \caption{Multiplicity distribution (left panel) and distribution of the eccentricity of each largest body in the system (right panel). Each distribution is colour-coded based on the total mass bins: low-mass, mid-mass, and high-mass (see text for more details). Dashed line marks the eccentricity value of e = 0.1 (see text for more details).}
    \label{fig:N_and_ecc_distrib}
\end{figure} 

We computed the total planetary mass for the subsample of multi-planetary systems, using measured values where available and lower limits otherwise, to assess its dependence on stellar mass and metallicity (Fig. \ref{fig:mtot_feh}). Our results indicate a clear positive correlation between total planetary mass and stellar mass across the parameter space considered. This finding is consistent with the established relationship between stellar mass and the dust and gas masses of protoplanetary discs \citep{Pascucci2016,Testi2022}, suggesting that the initial disc mass budget is a primary factor in setting the ultimate planetary system mass consistently with the results of population synthesis studies \citep[e.g.][]{Savvidou2023,Burn2025}. Upon dividing the sample by metallicity, a strong positive trend is evident for solar (-0.2 < \feh < 0.2) and super-solar stars (\feh > 0.2). We attribute this enhanced planetary mass growth to the greater abundance of solid materials in the protoplanetary discs allowed by the larger metallicity \citep[e.g.][]{Filomeno2024}. In such metal-rich environments, planetary core accretion proceeds efficiently, leading to an earlier onset of runaway gas accretion while the disc is still massive and capable of supplying substantial gaseous material to the growing planet. In contrast, the sub-solar bin shows a flat trend, with total masses below the threshold of giant planets (i.e., 0.2 \Mj\ as defined in \citealt{Magrini2022,Tsantaki_2025}). In these metal-poor systems, slower core formation delays the onset of gas accretion until the disc is depleted \citep[e.g.][]{Savvidou2023,Filomeno2024}, limiting growth and promoting the formation of multiple, similarly sized planets rather than a single dominant one.

To probe the intra-system mass distribution of the planetary systems, we adopted the metric 
M$_{L} = {M_p}/{M_{tot}}$ \citep{Chambers2001} that we expressed as the percentage $[\%]$ of the total planetary mass, M$_{tot}$, contained in the most massive planet in the system (of  mass M$_P$). This allowed us to analyse the architectural trends as a function of the stellar mass. In Fig. \ref{fig:mtot_mlarge} we see that, in the same M$_{\rm tot}$ range of the systems around stars with sub-solar metallicity discussed above (see also Fig. \ref{fig:mtot_feh}), the most massive planet typically contains 25-75\% of the total planetary mass, a range encompassing systems with relatively similar planets to systems with a mass concentration comparable to that of the Solar System (M$_{L,Jupiter}\approx0.7$).
Systems with a higher mass concentration (M$_{L} > 75\%$) exclusively have a multiplicity of N  = 2 and are among the most massive systems in the sample (the majority is above M$_{\rm tot} > 0.2$ \Mj). This suggests that, for high-mass systems, either dynamical instability or efficient accretion by a single body shaped their final architecture. Conversely, systems with a lower mass concentration  (M$_{L} < 50\%$) are characterised by $N >2$. All of them have M$_{\rm tot} < 0.2$\Mj, and show an opposite trend, with M$_{\rm tot}$ decreasing with the stellar mass.
\begin{figure}[t]
    \centering
   \includegraphics[width=0.95\linewidth]{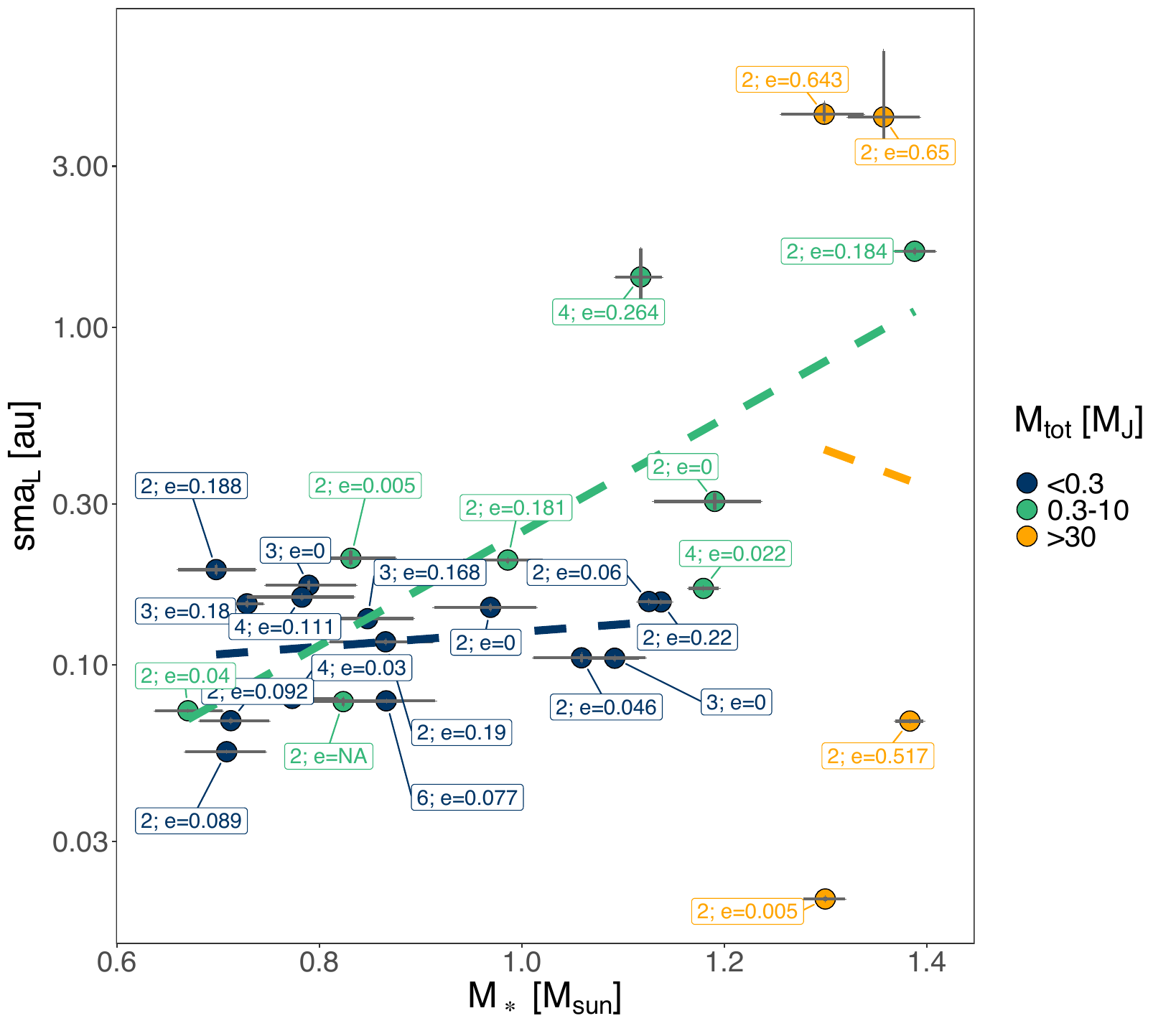}
    \caption{Semi-major axis of the largest body in the system, as a function of the host stellar mass, and colour-coded by the total mass in each system. The dashed regression lines are plotted for trend identification purposes. Labels indicate the multiplicity, N, and the eccentricity for each planet. Stellar mass and semi-major axis errors are reported for each system.}
    \label{fig:sma_mass}
\end{figure} 

To further investigate these behaviours, we divided the sample into three total-mass bins: low-mass systems (M$_{tot} < 0.2$ \Mj), mid-mass systems ($ 0.2 < M_{tot} < 10$ \Mj), and high-mass systems ($M_{tot} > 10$ \Mj). High-mass systems predominantly have N = 2 (Fig. \ref{fig:N_and_ecc_distrib}, left). This suggests either the formation of fewer, massive planets, or a later dynamical evolution via planet-planet scattering, which can create a dominant body through collisions or ejections \citep{Zinzi2017,Turrini2020}. Mid-mass systems show a binomial multiplicity distribution (peaks at N=2 and N=4), while low-mass systems span a wider range (N up to 6), consistent with Kepler's ‘peas-in-a-pod’ pattern \citep{Weiss2018}.

To probe their dynamical histories, we analysed the eccentricity of the largest planet in each system (Fig. \ref{fig:N_and_ecc_distrib}, right). High-mass systems are dominated by planets with $e > 0.5$, strongly indicating a chaotic, scattering-driven evolution history \citep{Zinzi2017,Turrini2022}. Mid-mass systems have a binomial eccentricity distribution split near $e = 0.1$, likely separating dynamically evolved from primordial systems. Low-mass systems show a more continuous eccentricity distribution, suggesting only moderate instabilities occurred, as weaker mutual perturbations are less likely to be catastrophic \citep{Turrini2020,Turrini2022}.

Finally, we examined the orbital separation of the largest planets (Fig. \ref{fig:sma_mass}). For low-mass systems, the semi-major axis shows little scatter. In contrast, the dispersion increases dramatically for mid- and high-mass systems, with the most massive planets spanning a wide range of orbital distances (0.02–3 AU). This broad distribution further supports once again a history of chaotic evolution for the most massive planetary systems resulting in wider orbital architectures \citep{Rasio1996,Weidenschilling1996}.

\section{Conclusions}\label{sec:summary}
We have presented analysis of 18 high-temperature 
(\teff\,$>6800$\,K) stars in the Ariel candidate sample
using an internally consistent methodology optimised for hot, rapidly rotating stars. 
The results complement the FGK host-star studies of \citet{Magrini2022} and \citet{Tsantaki_2025}, 
extending the stellar parameter space towards earlier spectral types. 
\begin{itemize}
\item{Stellar parameters:} We derived fundamental parameters (\teff, \logg, mass, \feh, \meh, \vmic, and $v\sin i$, ) for 18 stars using an iterative spectro–trigonometric method tailored for hot and fast-rotating stars.
The approach incorporates updated line lists, empirical oscillator-strength corrections, Balmer-line constraints, and large wavelength windows suitable for rapid rotators.
A comparison with 23 benchmark stars shows good overall agreement with previous Ariel analyses, ensuring that our results are internally consistent and placed on a comparable scale with the cooler-star studies in the Ariel MCS.
\item{Chemically peculiar stars:}
Several targets exhibit properties providing a preliminary indication of possible Am-type chemical peculiarity. A subset of the sample (HATS-70, KELT-17, TOI-1431, and WASP-178) has already been identified as chemically peculiar in previous studies, while for the remaining stars the indications are based on rotational properties and global metallicity trends. However, a detailed element-by-element abundance analysis is required to confirm the presence and degree of chemical peculiarity for the full sample, and will be carried out in future work.
\item{Kinematics:} The entire hot-star sample is consistent with thin-disc kinematics, in line with their relatively young ages and near-solar metallicities.
\item{Planet-star correlations:} The trends previously observed for FGK hosts also persist for hotter stars. Giant-planet radii increase toward higher stellar masses, primarily reflecting stronger stellar irradiation, while remaining metallicity dependent: at a given stellar mass, planets around metal-poor hosts tend to have larger radii, while planets around metal-rich hosts are generally more compact, consistent with differences in heavy-element content. The density--stellar-mass plane shows the same qualitative behaviour, although with additional scatter caused in part by heterogeneous planetary parameters and mass upper limits, which can artificially affect the inferred densities.
\end{itemize}
Overall, our results indicate that the star–planet correlations identified for FGK hosts  also extend to more massive, early-type stars. These findings highlight the importance of including high-temperature stars when assessing the diversity of planets
and refining the target selection for the Ariel mission. Furthermore, they highlight how stellar properties shape the global architecture of multi-planet systems and the path of their dynamical evolution, imprinting on their multiplicity, mass and mass distribution, and dynamical excitation.

\section{Data availability} 
All parameters presented here (Table~\ref{tab:Table_hot_results}) are available via both the CDS\footnote{\url{http://cdsportal.u-strasbg.fr}} and the Ariel Stellar Catalogue\footnote{\url{https://bit.ly/ArielStellarCatalogue}}. The latter also includes atmospheric parameters (\feh, \teff, \logg, \vmic , \vsini), kinematic properties and C, N, O  abundances for the cooler sample of planet-host FGK dwarf stars.

\begin{acknowledgements}
This work has been developed within the framework of the Ariel ``Stellar Characterisation'' and ``Planet Formation'' working groups of the ESA Ariel space mission Consortium. 
H.R., S.B., C.P.F., A.L., V.M. and M.K. acknowledge funding from the European Union's Horizon Europe research and innovation programme under grant agreement No. 101079231 (EXOHOST) and from UK Research and Innovation (UKRI) under the UK government’s Horizon Europe funding guarantee (grant number 10051045). The research was conducted using the ESA Estonia research infrastructure funded by the Estonian Research Council  grant TARISTU24-TK3. This work has made use of the VALD database, operated at Uppsala University, the Institute of Astronomy RAS in Moscow, and the University of Vienna; the VizieR catalogue access tool CDS, Strasbourg, France (DOI : 10.26093/cds/vizier). 
C.D. acknowledges financial support from the grant RYC2023-044903-I, funded by MCIU/AEI/10.13039/501100011033 and by the ESF+, and from the INAF initiative ``IAF Astronomy Fellowships in Italy'', grant name \textit{GExoLife}. 
D.T. acknowledges support from the Italian Space Agency (ASI) through the ASI-INAF grant no. 2021-5-HH.0 plus addenda no. 2021-5-HH.1-2022 and 2021-5-HH.2-2024, the COST Action CA22133 PLANETS, and the European Research Council via the Horizon 2020 Framework Programme ERC Synergy “ECOGAL” Project (project ID GA-855130).
D.B. acknowledges funding support by the Italian Ministerial Grant PRIN 2022, ``Radiative opacities for astrophysical applications'', no. 2022NEXMP8, CUP C53D23001220006. 
L.M. thank INAF for the support (Large Grants EPOCH and WST), the Mini-Grants Checs (1.05.23.04.02), and the financial support under the National Recovery and Resilience Plan (NRRP), Mission 4, Component 2, Investment 1.1, Call for tender No. 104 published on 2.2.2022 by the Italian Ministry of University and Research (MUR), funded by the European Union – NextGenerationEU – Project ‘Cosmic POT’ Grant Assignment Decree No. 2022X4TM3H by the Italian Ministry of the University and Research (MUR).

K.G.H. acknowledges support from NCN grant 2023/49/B/ST9/01671.
H.R. thanks Luca Fossati, Anish Amarsi, and Indrek Kolka for their support and insightful comments.
The team is very grateful to the service astronomers who performed our observations at ESO TNG (with HARPS-N during A41, A46). 
Based on observations collected at the European Southern Observatory under ESO programmes and with the SOPHIE spectrograph on the 1.93 m telescope at the Observatoire de Haute-Provence (CNRS), France, by the SOPHIE RPE Consortium (program PNP.CONS).

 \end{acknowledgements}

\bibliographystyle{aa}
\bibliography{Ariel_bib}

\newpage

\appendix  

\section{Line list and  oscillator strength corrections}\label{Oscillator_strength_corrections} 

\begin{figure*}[!ht]
    \centering
    \begin{subfigure}
        \centering
    \includegraphics[width=0.99\linewidth]{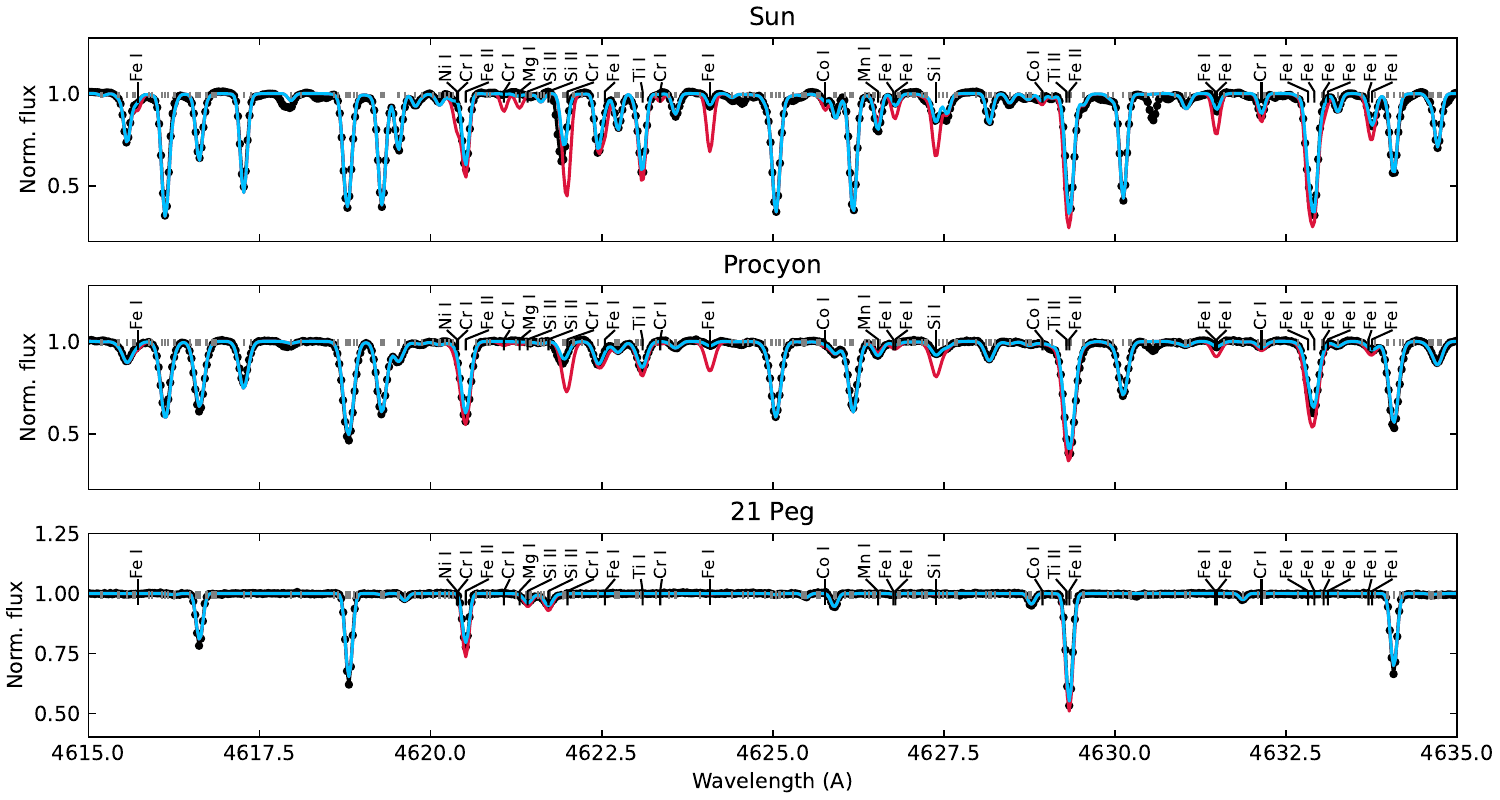}
    \end{subfigure}

        \begin{subfigure}
        \centering
    \includegraphics[width=0.99\linewidth]{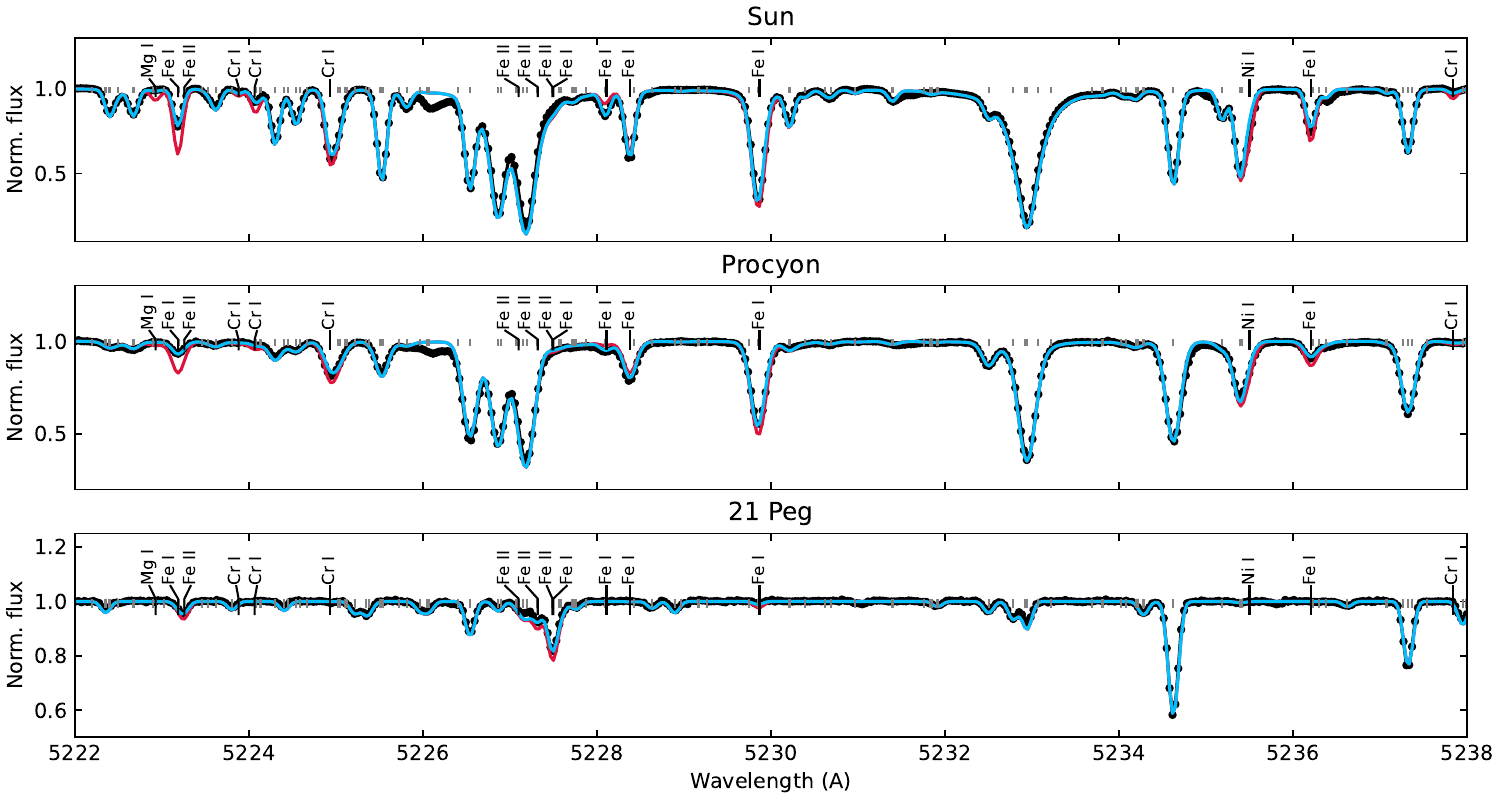}
    \caption{Examples of $\log gf$ corrections calibrated using the Sun, Procyon, and 21 Peg. The black line shows the observed spectrum, the red line the synthetic spectrum computed with the original VALD3 $\log gf$ values, and the blue line the final corrected model. Only lines for which $\log gf$ values were adjusted are labelled, while small ticks mark the positions of unmodified lines.}
    \label{fig:loggf}
    \end{subfigure}
\end{figure*}

New extractions from the VALD3 database were made (using version 3735M, in May 2024). Extractions were made for lines between 4000 and 10000 \AA, using the `extract stellar' tool, for \teff\ of 4500, 5000, 6000, 7000, 8000, 9000, and 10000 K, at logg of 4.5 and solar abundances, a depth threshold of 0.01, and using the `extended van der Waals broadening' (for lines with coefficients in the Anstee, Barklem, \& O'Mara formalism). This provided us with a set of seven VALD line lists, valid for stars from early A to late K-types. 

The seven VALD line lists were merged with each other, and the combined VALD line list was merged with the \citet{Heiter2021} `Y' and `U' lines. At both stages in merging, care was taken to avoid introducing duplicate lines. Small discrepancies in wavelengths and excitation energies from different sources were accounted for. When these differences were found, we checked the J quantum numbers, electron configuration, and term symbol strings. Flags for isotopic and hyperfine splitting components were also checked. In several cases the electron configuration or term symbol strings were incomplete or corrupted. In those cases, the possible and duplicate lines were evaluated by comparing synthetic spectra to observations of reference stars, and if including both candidate transitions better reproduced the observation then we concluded they were two different real lines and both were retained. Otherwise the single line that better reproduced the observation was retained. When duplicates between \citet{Heiter2021} `Y' lines and VALD were found, the `Y' lines were used. When duplicates were found between VALD line lists, lines were selected based on VALD's default selection weights, if the lines came from different sources.

Empirical oscillator strength ($\log gf$) corrections were derived for this line list by comparing synthetic and observed spectra of three standard stars: the Sun, Procyon, and 21 Peg. The observations were obtained with the ESPaDOnS spectrograph at the Canada France Hawaii Telescope, and the solar observation was obtained as reflected from the moon. Parameters for Procyon and the Sun were taken from \citet{Heiter2015}, and parameters for 21 Peg were taken from \citet{Fossati2009} and \citet{Mashonkina2020}. When a correction was derived for one star, it was checked against the other stars, to ensure it did not degrade the quality of the fits to those observations. In rare cases where a significant disagreement between stars was found (e.g. due to an unidentified line blend, or temperature dependent non-LTE effects), a manual assessment was made to verify the quality of observations and lack of apparent blends, and then a compromise value was adopted to minimise the total discrepancy between models and observations.  
We emphasise that these empirical corrections do not necessarily better represent the true $\log gf$ values, since they are influenced by, and partially compensate for, limitations of the synthetic spectra, such as the assumption of LTE.

Lines for correction were identified manually, by comparing synthetic spectra to observations, and looking for discrepancies greater than 5\% of the continuum, where the problematic line could be distinguished. In cases where the origin of the problem was ambiguous, either due to line blending or due to an apparently missing line in the list, no correction was made. Corrections were derived by fitting synthetic spectra to observations by $\chi^2$ minimisation, with $\log gf$ as a free parameter. Lines requiring corrections were also checked against the NIST Atomic Spectra Database \citep{NIST_ASD}, and if a high quality $\log gf$ was present, this was used instead of an empirical correction. A sample of corrections in two different wavelength regions are shown in Fig. \ref{fig:loggf}.  
This manual comparison of models to observations was also used to determine which \citet{Heiter2021} `U' lines produced superior fits than VALD data, and so which were retained. Empirical corrections were  not made for the highest quality `Y' lines.

\newpage

\section{Supplementary Material}

\begin{figure*}[!hbpt]
    \centering
    \includegraphics[width=1\linewidth]{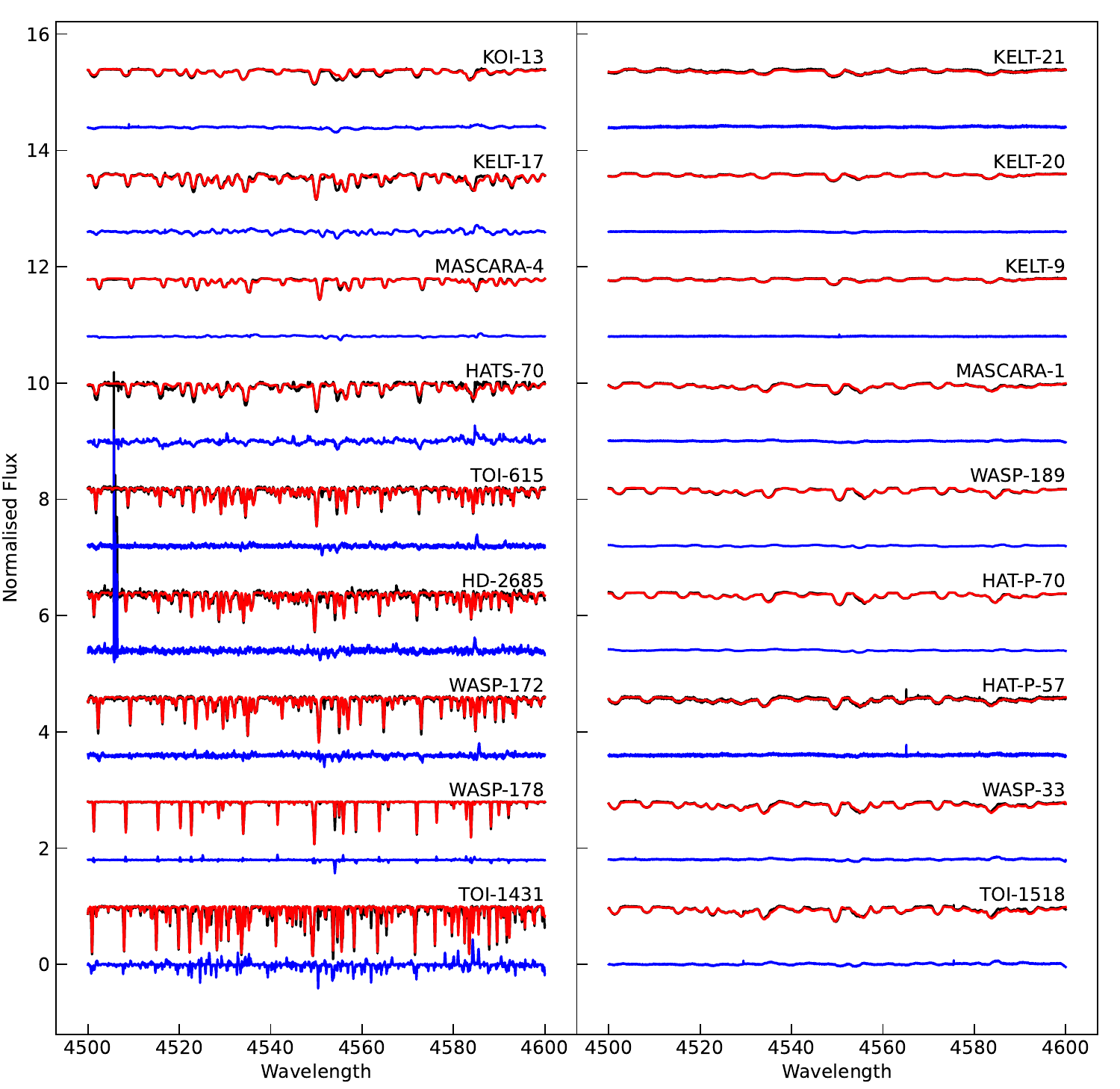}
    \caption{Observed, synthetic, and difference spectra (black, red, blue lines)
    for stars in the sample, sorted by \vsini\ from left bottom to top right.}
    \label{fig:obs-synt-diff}
\end{figure*}

\onecolumn

\clearpage  

\begin{sidewaystable*}
\caption{Basic parameters and notes for the exoplanets and companions in our sample.}
\label{tab:hj-systems}
\centering
\scriptsize
\setlength{\tabcolsep}{3pt}
\begin{tabular}{lccccp{4cm}p{6cm}p{3.3cm}}
\hline\hline
Planet &
$ a $ (AU) &
Mass (M$_{\rm Jup}$) &
Radius (R$_{\rm Jup}$) &
$T_{\rm eq}$ (K) &
Companion(s) &
Notes &
References \\
\hline
HAT-P-57\,A\,b &
$0.0406 \pm 0.0011$ &
$\leq 1.85$ &
$1.413 \pm 0.054$ &
$2200 \pm 76$ &
HAT-P-57\,B ($0.61 \pm 0.10\,M_\odot$) and HAT-P-57\,C is ($0.53 \pm 0.08\,M_\odot$); projected separation B–C is $68$ AU; distance from A $800$ AU. &
HAT-P-57\,C may be a white dwarf. HAT-P-57\,A\,b has featureless spectra. Stellar pulsations in HAT-P-57\,A can obscure or mimic the planetary signal, so its planetary nature has been debated. &
\citep{Hartman2015,Stranget2022,Jiang2023} \\
HAT-P-70\,b &
$0.04739 \pm 0.00106$ &
$\leq6.78 _{-6.77}^{+0}$ &
$1.66 \pm 0.20$ &
$2562 \pm 43$ &
– &
Atmospheric species detected include Ca, Fe, Mg, Ni, Cr, Mn, Na and V. A dominant subsolar–antisolar wind circulation pattern has been suggested. &
\citep{Zhou_2019_HAT-P-70,Gandhi2023,Langeveld2025} \\
HATS-70\,b &
$0.0363 \pm 0.0007$ &
$12.9 \pm 1.8$ &
$1.38 \pm 0.08$ &
$2730 \pm 160$ &
- &
HATS-70\,b is more likely a brown dwarf than a planet. &
\citep{Zhou2019a} \\
HD\,2685\,b &
$0.0568 \pm 0.0006$ &
$1.17 \pm 0.12$ &
$1.44 \pm 0.05$ &
$2061$ &
– &
The coolest host star in the sample. &
\citep{Jones2019} \\
KELT-17\,b &
$0.04881 \pm 0.00065$ &
$1.31 \pm 0.29$ &
$1.525 \pm 0.065$ &
$2087$ &
– &
KELT-17 is an Am-type star. The planetary orbit is strongly misaligned (projected spin–orbit angle $\lambda \sim 84^\circ$). &
\citep{Garai2022,zhou2016,Hansen2012} \\
KELT-20\,A\,b &
$0.0542 \pm 0.0021$ &
$\leq3.382$ &
$1.741 \pm 0.074$ &
$2252 \pm 78$ &
KELT-20\,B, possibly a brown dwarf, at a projected separation of $6595$ AU. &
The system exhibits a near-perfect spin–orbit alignment ($3.4^\circ \pm 2.1^\circ$). Supersonic day–night winds are inferred in the planetary atmosphere. &
\citep{lund2017,Finnerty2025,Eeles-Nolle2025} \\
KELT-21\,A\,b &
$0.05224 \pm 0.0016$ &
$\leq3.91$ &
$1.586 \pm 0.04$ &
$2025 \pm 30$ &
KELT-21\,B and KELT-21\,C, mid-M dwarfs; separation B–C is $20$ AU; distance from A is $\sim 500$ AU. &
KELT-21\,A\,b has a well-aligned orbit; Kozai–Lidov oscillations are therefore unlikely to have driven the migration. &
\citep{Johnson2018,Hamers2017,Saha2024} \\
KELT-9\,A\,b &
$0.03462 \pm 0.00078$ &
$2.88 \pm 0.84$ &
$1.891 \pm 0.061$ &
$3921 \pm 182$ &
KELT-9\,B, an M dwarf at a projected separation of $2671$ AU. &
The mass-loss rate is approximately $10^{12.8 \pm 0.3}$ g s$^{-1}$; the planet could lose its atmosphere within $\sim 600$ Myr. &
\citep{Stassun2019,Mugrauer2019,Michel2024,borsa2019,Wyttenbach2020,gaudi2017,Harre2023} \\
KOI-13\,A\,b &
$0.03641 \pm 0.001$ &
$\leq 9.28$ &
$1.512 \pm 0.035$ &
$2550 \pm 80$ &
KOI-13\,B (A-type star, $1.69\,M_\odot$, orbital period $\sim 1000$ yr) and KOI-13\,C (low-mass K dwarf, $0.7\,M_\odot$, orbiting B with $P=65.8$ d); projected separation from A is $569$ AU. &
The planet has an oblique orbit. The system shows a strong spin–orbit misalignment and a varying transit duration, probably caused by nodal precession. &
\citep{Szabo2011,Szabo2020,Santerne2012,Barnes2011,Esteves2015,Szabo2014} \\
MASCARA-1\,b &
$0.043 \pm 0.005$ &
$3.7 \pm 0.9$ &
$1.597 \pm 0.3$ &
– &
– &
The planet has strongly misaligned orbit, $72.1^\circ \pm 2.5^\circ$. &
\citep{Hooton2022,Talens2017} \\
MASCARA-4\,A\,b &
$0.0474 \pm 0.004$ &
$1.675 \pm 0.241$ &
$1.515 \pm 0.07$ &
$2455 \pm 41$ &
MASCARA-4\,B, a late K/M dwarf at a projected separation of $744$ AU. &
The planet has very highly misaligned, retrograde orbit ($\lambda=250.34^\circ\pm0.14^\circ$). Rubidium and samarium have been detected in its atmosphere. &
\citep{Dorval2020,Michel2021,Zhang2022A,Saha2024,Jiang2023} \\
TOI-1431\,b &
$0.046 \pm 0.002$ &
$3.12 \pm 0.18$ &
$1.49 \pm 0.05$ &
$2370 \pm 70$ &
? &
TOI-1431 is an Am-type star. No atmosphere has yet been detected for the planet. &
\citep{Addison2021,Stangret2021} \\
TOI-1518\,b &
$0.0389 \pm 0.0011$ &
$\leq2.3$ &
$1.875 \pm 0.053$ &
$2892 \pm 38$ &
– &
The orbit undergoes nodal precession; transits are predicted to cease by $2194 \pm 70$ AD. &
\citep{Cabot2021,Watanabe2024} \\
TOI-615\,b &
$0.0678 \pm 0.0031$ &
$0.435 \pm 0.086$ &
$1.693 \pm 0.057$ &
$1666 \pm 24$ &
– &
One of the lowest-density exoplanets discovered to date. &
\citep{Psaridi2023} \\
WASP-172\,b &
$0.0694 \pm 0.001$ &
$0.47 \pm 0.1$ &
$1.57 \pm 0.1$ &
$1740 \pm 60$ &
– &
Sodium has been detected in its atmosphere. &
\citep{Hellier2019,Seidel2023} \\
WASP-178\,b &
$0.0558 \pm 0.001$ &
$1.66 \pm 0.12$ &
$1.81 \pm 0.09$ &
$2402 \pm 130$ &
– &
WASP-178 is a slowly rotating weak Am star. The planet’s atmosphere appears carbon-depleted and shows titanium depletion. &
\citep{Fossati_2025,Hellier2019b,Rodriguez2020,Lothringer2025} \\
WASP-189\,A\,b &
$0.05053 \pm 0.00098$ &
$1.99 \pm 0.16$ &
$1.619 \pm 0.021$ &
$3353 \pm 34$ &
WASP-189\,B, an M dwarf at a projected separation of $942$ AU. &
Upper-atmosphere studies indicate Mg\,II and possibly Fe\,II, with evidence for escaping magnesium and temperatures around $1.5\times10^4$ K. &
\citep{Lendl2020,Sreejith2023,Eeles-Nolle2025} \\
WASP-33\,A\,b &
$0.0239 \pm 0.00063$ &
$2.093 \pm 0.139$ &
$1.593 \pm 0.074$ &
$2781.70 \pm 41.10$ &
WASP-33\,B (M dwarf, 238 AU) and WASP-33\,C (G dwarf, 5955 AU). &
WASP-33\,A is a non-radial $\delta$\,Scuti pulsator (a $\gamma$\,Doradus–$\delta$\,Scuti hybrid candidate). The planet has a retrograde, nearly polar orbit ($108.19\pm 0.97$) and exhibits poor day–night heat redistribution. &
\citep{Chakrabarty2019a,Collier2010,VonEssen2020,Stassun2019,Mugrauer2019,vonEssen2014,Smith2011} \\
\hline
\end{tabular}
\tablefoot{$a$ is the orbital distance; $T_{\rm eq}$ is the zero-albedo equilibrium temperature. Stellar and planetary parameters are adopted from the cited discovery and follow-up papers.}
\end{sidewaystable*}

\onecolumn

\begin{table*}[ht]
\renewcommand{\arraystretch}{1.3}
\small
\centering
\caption{Fundamental parameters of benchmark stars analysed in this paper.}
\begin{tabular}{lllccccccc}
\hline\hline
ID & \multicolumn{1}{c}{$T_{\rm eff}$ [K]}& \multicolumn{1}{c}{$T_{\rm eff,\,Balmer}$ [K]} & $\log g$ [dex] & Mass [$M_\odot$] & Radius [$R_\odot$] & [Fe/H] & [M/H] & $\xi$ [km/s] & $v \sin i$ [km/s] \\
\hline
HAT-P-2 & 6609 ± 36 & 6609 ± 64 & 4.19 ± 0.02 & ${1.43}_{-0.02}^{+0.03}$ & ${1.61}_{-0.02}^{+0.03}$ & 0.18 ± 0.05 & 0.18 ± 0.01 & 1.7 ± 0.1 & 21.3 ± 0.0 \\
HAT-P-67 & 6541 ± 77 & 6562 ± 48 & 3.84 ± 0.05 & ${1.71}_{-0.04}^{+0.03}$ & ${2.64}_{-0.07}^{+0.08}$ & 0.14 ± 0.05 & 0.14 ± 0.08 & 1.7 ± 0.2 & 35.6 ± 0.6 \\
HD106315 & 6581 ± 77 & 6571 ± 57 & 4.35 ± 0.05 & ${1.22}_{-0.04}^{+0.05}$ & ${1.24}_{-0.03}^{+0.03}$ & -0.07 ± 0.08 & -0.05 ± 0.04 & 1.5 ± 0.1 & 13.4 ± 0.1 \\
HR858 & 6451 ± 31 & 6562 ± 48 & 4.36 ± 0.02 & ${1.24}_{-0.02}^{+0.03}$ & ${1.23}_{-0.02}^{+0.03}$ & 0.05 ± 0.04 & 0.05 ± 0.02 & 1.4 ± 0.1 & 7.6 ± 0.2 \\
K2-99 & 6215 ± 48 & 6247 ± 48 & 3.85 ± 0.05 & ${1.61}_{-0.09}^{+0.03}$ & ${2.51}_{-0.06}^{+0.08}$ & 0.16 ± 0.05 & 0.18 ± 0.01 & 1.7 ± 0.1 & 9.6 ± 0.0 \\
KELT-1 & 6696 ± 72 & 6723 ± 91 & 4.25 ± 0.05 & ${1.41}_{-0.06}^{+0.04}$ & ${1.48}_{-0.04}^{+0.05}$ & 0.16 ± 0.10 & 0.17 ± 0.04 & 1.8 ± 0.1 & 51.5 ± 0.1 \\
KELT-24 & 6683 ± 45 & 6712 ± 141 & 4.29 ± 0.03 & ${1.44}_{-0.02}^{+0.03}$ & ${1.44}_{-0.02}^{+0.02}$ & 0.31 ± 0.07 & 0.32 ± 0.03 & 1.6 ± 0.1 & 21.6 ± 0.2 \\
KELT-7 & 6856 ± 113 & 6951 ± 162 & 4.16 ± 0.06 & ${1.61}_{-0.08}^{+0.04}$ & ${1.75}_{-0.07}^{+0.08}$ & 0.38 ± 0.12 & 0.40 ± 0.02 & 2.1 ± 0.1 & 70.8 ± 0.4 \\
KELT2A & 6344 ± 38 & 6440 ± 32 & 4.07 ± 0.03 & ${1.45}_{-0.02}^{+0.03}$ & ${1.86}_{-0.03}^{+0.04}$ & 0.18 ± 0.04 & 0.18 ± 0.01 & 1.4 ± 0.1 & 8.0 ± 0.1 \\
KOI-12 & 6922 ± 55 & 6910 ± 82 & 4.27 ± 0.03 & ${1.46}_{-0.02}^{+0.02}$ & ${1.47}_{-0.03}^{+0.03}$ & 0.15 ± 0.04 & 0.12 ± 0.04 & 1.7 ± 0.2 & 67.1 ± 0.4 \\
TOI-677 & 6328 ± 36 & 6384 ± 100 & 4.33 ± 0.02 & ${1.22}_{-0.04}^{+0.03}$ & ${1.25}_{-0.03}^{+0.02}$ & 0.08 ± 0.03 & 0.09 ± 0.02 & 1.4 ± 0.1 & 6.5 ± 0.1 \\
V1298Tau & 5067 ± 101 & 5217 ± 92 & 4.10 ± 0.06 & ${0.78}_{-0.03}^{+0.03}$ & ${1.28}_{-0.04}^{+0.05}$ & 0.05 ± 0.11 & 0.09 ± 0.10 & 1.6 ± 0.3 & 27.1 ± 0.4 \\
WASP-120 & 6649 ± 44 & 6661 ± 105 & 4.18 ± 0.02 & ${1.46}_{-0.02}^{+0.01}$ & ${1.64}_{-0.03}^{+0.03}$ & 0.19 ± 0.02 & 0.18 ± 0.03 & 1.7 ± 0.1 & 15.7 ± 0.2 \\
WASP-129 & 5994 ± 141 & 6108 ± 16 & 4.36 ± 0.08 & ${1.15}_{-0.05}^{+0.04}$ & ${1.19}_{-0.05}^{+0.06}$ & 0.20 ± 0.09 & 0.22 ± 0.05 & 1.1 ± 0.1 & 3.8 ± 0.2 \\
WASP-166 & 6172 ± 28 & -- & 4.38 ± 0.02 & ${1.25}_{-0.02}^{+0.01}$ & ${1.21}_{-0.02}^{+0.01}$ & 0.28 ± 0.04 & 0.27 ± 0.01 & 1.3 ± 0.0 & 4.9 ± 0.3 \\
WASP-167 & 7303 ± 18 & 7175 ± 98 & 4.16 ± 0.02 & ${1.66}_{-0.02}^{+0.02}$ & ${1.79}_{-0.02}^{+0.03}$ & 0.12 ± 0.04 & 0.17 ± 0.03 & 2.7 ± 0.2 & 49.7 ± 0.1 \\
WASP-17 & 6778 ± 34 & 6785 ± 63 & 4.24 ± 0.03 & ${1.36}_{-0.02}^{+0.03}$ & ${1.48}_{-0.03}^{+0.03}$ & -0.01 ± 0.06 & 0.00 ± 0.06 & 1.7 ± 0.1 & 10.4 ± 0.1 \\
WASP-19 & 5568 ± 84 & 5681 ± 47 & 4.44 ± 0.06 & ${1.00}_{-0.03}^{+0.04}$ & ${1.01}_{-0.02}^{+0.03}$ & 0.26 ± 0.06 & 0.34 ± 0.04 & 1.1 ± 0.1 & 5.4 ± 0.2 \\
WASP-7 & 6600 ± 39 & 6574 ± 30 & 4.28 ± 0.02 & ${1.38}_{-0.02}^{+0.02}$ & ${1.42}_{-0.02}^{+0.02}$ & 0.20 ± 0.04 & 0.21 ± 0.01 & 1.6 ± 0.1 & 18.2 ± 0.2 \\
WASP-8 & 5659 ± 78 & 5755 ± 40 & 4.48 ± 0.05 & ${1.06}_{-0.03}^{+0.03}$ & ${1.01}_{-0.03}^{+0.03}$ & 0.30 ± 0.05 & 0.36 ± 0.02 & 1.2 ± 0.1 & 1.3 ± 0.6 \\
XO-3 & 7045 ± 44 & 7016 ± 43 & 4.30 ± 0.03 & ${1.43}_{-0.02}^{+0.02}$ & ${1.42}_{-0.03}^{+0.02}$ & 0.04 ± 0.04 & 0.08 ± 0.03 & 1.9 ± 0.1 & 19.5 ± 0.2 \\
XO-6 & 6816 ± 49 & 6734 ± 79 & 4.32 ± 0.03 & ${1.36}_{-0.03}^{+0.02}$ & ${1.35}_{-0.03}^{+0.02}$ & -0.01 ± 0.05 & -0.02 ± 0.03 & 1.5 ± 0.2 & 46.4 ± 0.6 \\
XO-7 & 6086 ± 70 & 6194 ± 77 & 4.23 ± 0.04 & ${1.30}_{-0.02}^{+0.02}$ & ${1.45}_{-0.04}^{+0.04}$ & 0.33 ± 0.03 & 0.34 ± 0.04 & 1.1 ± 0.0 & 6.4 ± 0.1 \\
\hline
\label{Table:BM_Results_Balmer}
\end{tabular}
\tablefoot{The fundamental parameters are: effective temperature derived from metal-line analysis (K), Balmer-line effective temperature (K), surface gravity (dex), stellar mass ($M_\odot$), stellar radius ($R_\odot$), iron abundance [Fe/H] (dex), overall metallicity [M/H] (dex), microturbulent velocity (km/s) and projected rotational velocity (km/s). All parameter values are reported with 1$\sigma$ uncertainties. \(T_{\rm eff,\,Balmer}\) values are derived from Balmer line fits excluding H$\alpha$ due to tellurics.}
 \label{BM_tabel}
\end{table*}

\twocolumn
\begin{figure}[h]
    \centering
    \includegraphics[width=1\linewidth]{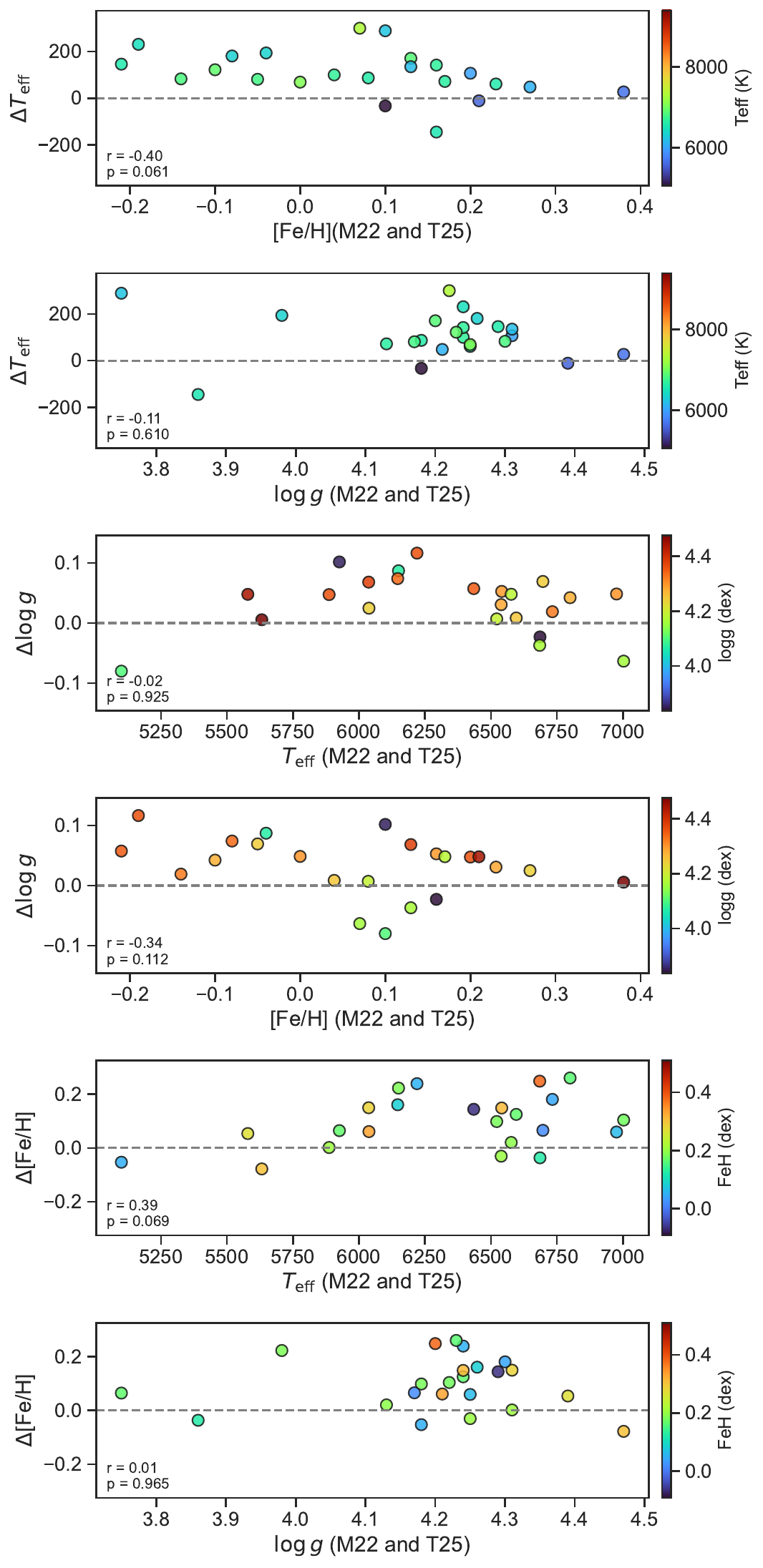}
    \caption{Residuals between our this work and literature values from \citet{Magrini2022} and \citet{Tsantaki_2025} as a function of the literature values. Points are colour-coded by our derived value of the parameter shown on the y-axis. Pearson correlation coefficients ($r$) and $p$-values are annotated in each subplot. }
    \label{fig:Correl_BM_param}
\end{figure}

\begin{figure}[ht!]
    \centering

    \begin{subfigure}
        \centering
        \includegraphics[width=1\linewidth,trim=0.4cm 0.8cm 0.3cm 0.30cm]{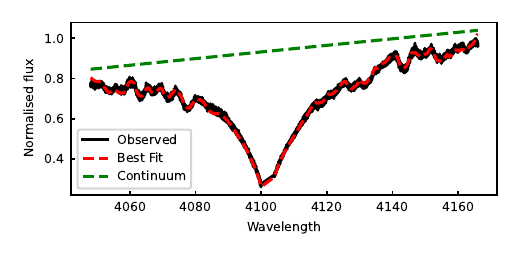}
    \end{subfigure}

    \begin{subfigure}
        \centering
        \includegraphics[width=1\linewidth,trim=0.4cm 0.8cm 0.3cm 0.25cm]{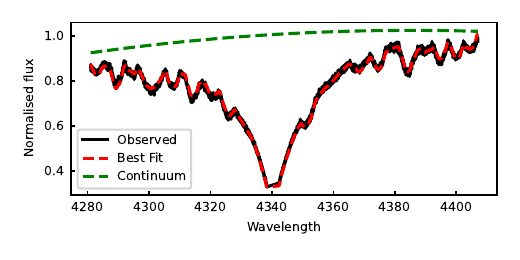}
    \end{subfigure}

    \begin{subfigure}
        \centering
        \includegraphics[width=1\linewidth,trim=0.4cm 0.8cm 0.3cm 0.25cm]{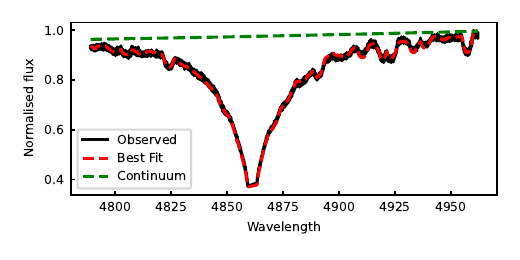}
        \label{fig:KELT21_beta}
    \end{subfigure}
    \caption{Simultaneous continuum and \teff\ Balmer line fits for KELT-21. The continuum is modelled as a second-degree polynomial. From top to bottom, the panels show H$\delta$, H$\gamma$ and H$\beta$.}
    \label{fig:KELT21_balmer}
\end{figure}

\begin{figure}
    \centering
    \includegraphics[width=1\linewidth]{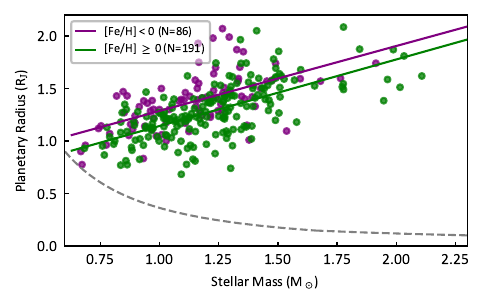}
    \caption{Planetary radius as a function of stellar mass for giant planets 
($M_{\mathrm{p}}>0.2\,M_{\mathrm{J}}$ and $R_{\mathrm{p}}>0.6\,R_{\mathrm{J}}$),  
split by stellar metallicity.  
Purple and green points show planets orbiting metal-poor (\feh $<0$)  
and metal-rich (\feh $\ge 0$) stars, respectively; solid lines indicate  
the corresponding linear fits.  
The grey dashed curve illustrates the qualitative scaling of a fixed transit 
depth with stellar mass (assuming $R_\star \propto M_\star^\alpha$).  
This curve is intended to show the decreasing detectability of small planets 
around more massive stars and does not represent a strict survey 
detection limit.}
    \label{fig:r_vs_m_2bins}
\end{figure}

\begin{figure*}
    \centering
    \includegraphics[width=1\linewidth]{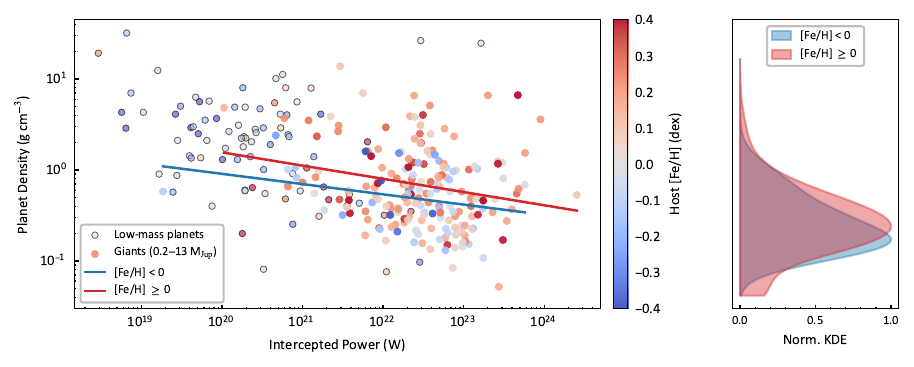}
    \caption{Planet density as a function of intercepted stellar power.
        Filled circles mark low-mass planets (not included in the fit), while outlined symbols indicate the high-mass planet subset (\(0.2 < M_{\rm p} < 13\,M_{\rm Jup}\)) used for the power-law regressions. Points are colour-coded by host-star metallicity. 
        Blue and red lines represent the best-fit relations for high-mass planets orbiting metal-poor (\([\mathrm{Fe/H}]<0\)) and metal-rich (\([\mathrm{Fe/H}]\ge0\)) hosts, respectively.
        \textit{Right:} Normalised kernel-density distributions of the giant-planet densities for the two metallicity bins.}
        
    \label{fig:pl_density_Pinc}
\end{figure*}

\begin{figure}[!htb]
    \centering
    \includegraphics[width=0.9\linewidth,trim=5cm 0.0cm 5cm 0.0cm]{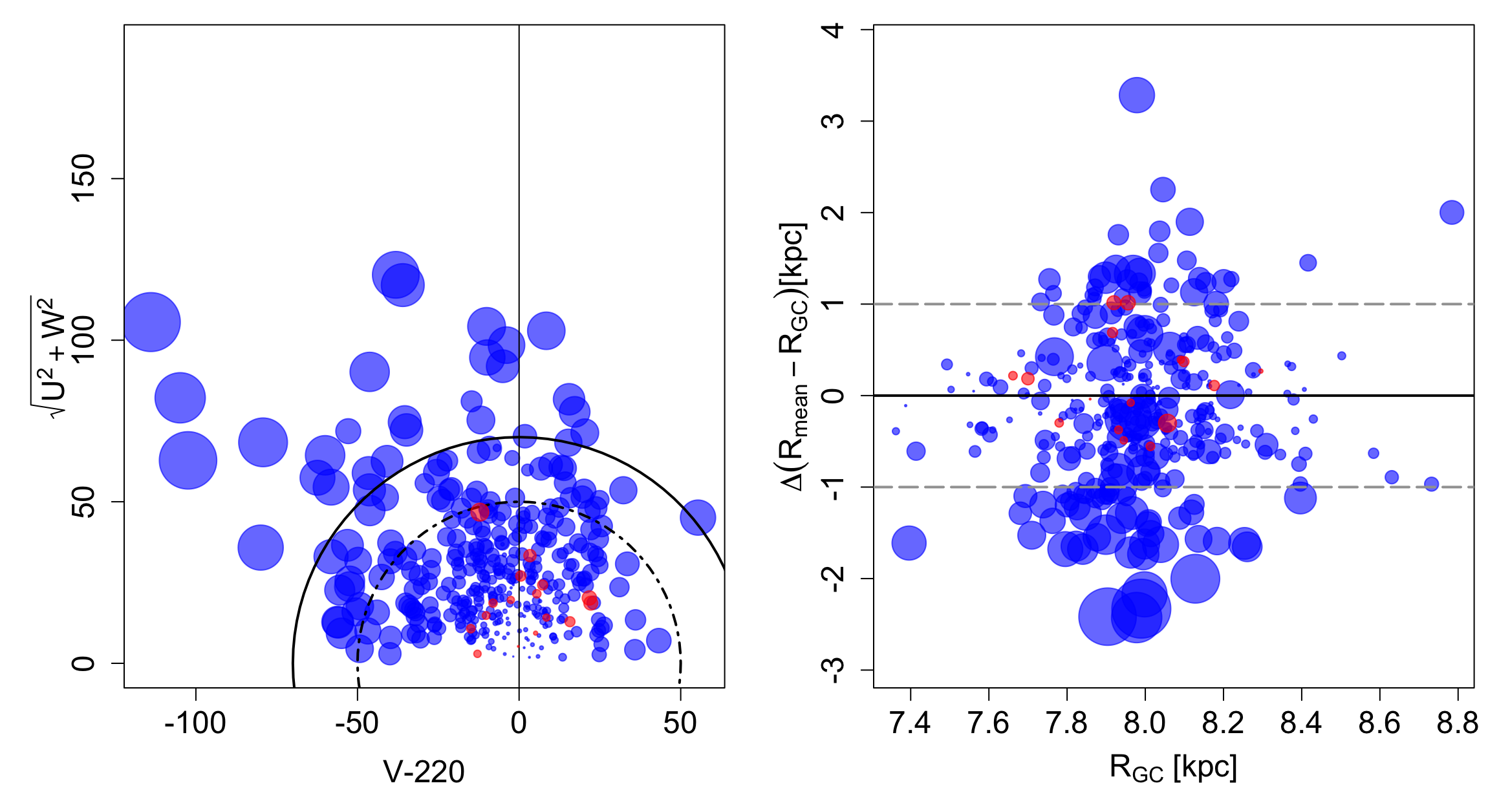}
    \caption{\textit{Left:} Toomre diagram with dashed and solid lines indicating $|V_{\rm tot}|=50$ and $70$\,\kms, respectively, used to separate thin disc, thick disc, and intermediate populations. \textit{Right:} Orbital parameter space showing the difference between mean and current Galactocentric radius ($R_{\rm mean}-R_{\rm GC}$) as a function of $R_{\rm GC}$. The hot star sample is shown in red and the T25 comparison sample in blue; the symbol size scales with orbital eccentricity.}
    \label{fig:Toomre}
\end{figure}

\begin{figure}[h]
    \centering
     \includegraphics[width=\linewidth]{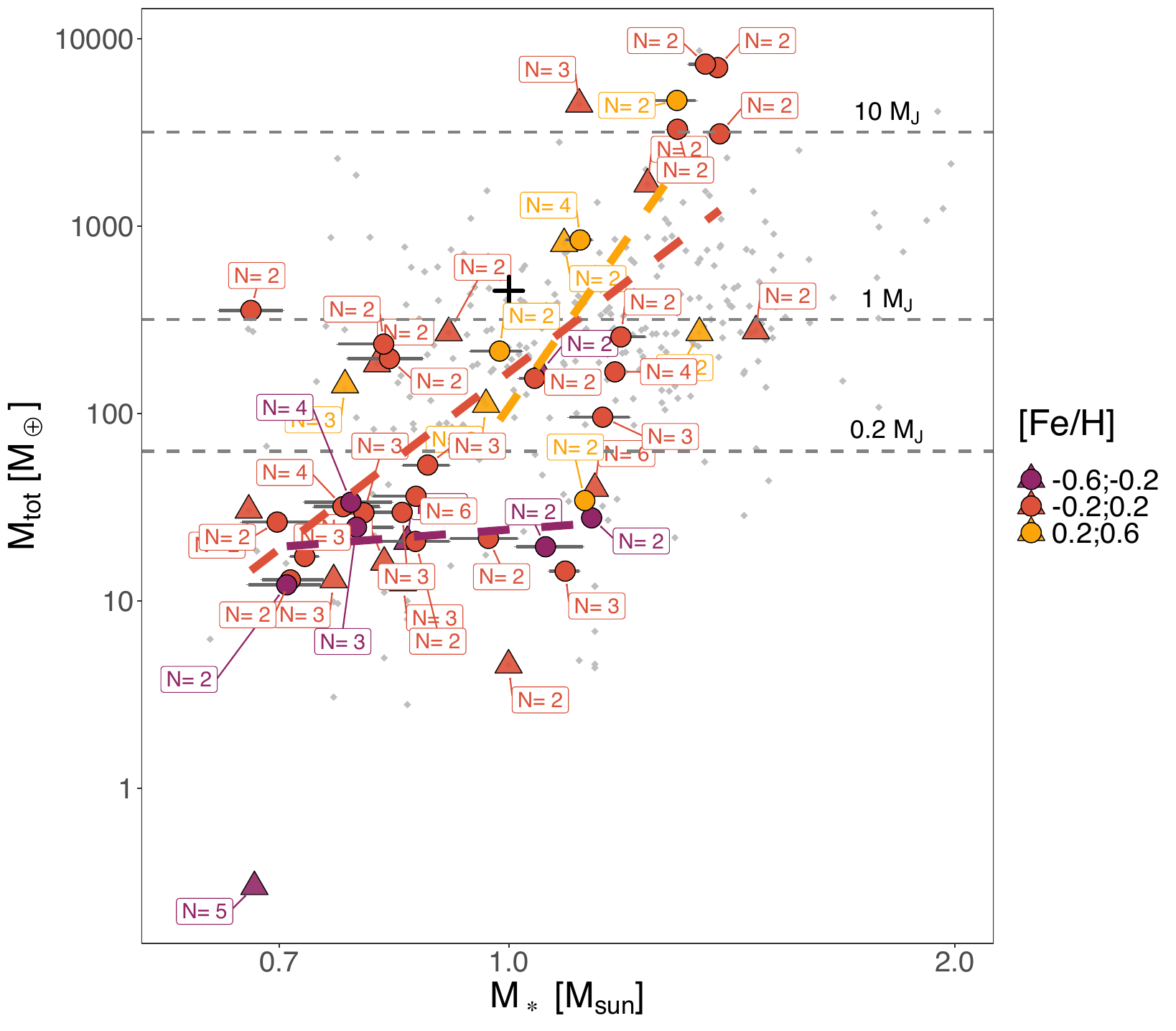}
     \caption{Total planetary mass available in each planetary system (circle), as a function of the stellar mass, colour-coded by the stellar \feh.  Triangles represents those systems for which we could only determine a lower limit of the total mass. Gray dots represents the mass of planets in single planet systems. Dashed lines are plotted to provide a visual range of masses in term of \Mj. The black cross identifies the Solar System and stellar mass errors are reported for each system.}
    \label{fig:mtot_feh}
\end{figure}

\clearpage
\end{document}